\documentclass[aps,pra,onecolumn,superscriptaddress,floatfix,nofootinbib,showkeys,]{revtex4-2}
\usepackage{amsmath,amssymb,graphicx,siunitx,hyperref}
\usepackage[english]{babel}

\begin{document}

\title{Experimental investigation of the effect of dispersion on squeezing generation in a synchronously pumped optical parametric oscillator}

\author{Edoardo~Suerra}
\email{edoardo.suerra@unimi.it}
\affiliation{Dipartimento di Fisica, Università degli Studi di Milano, Via Celoria 16, Milan, Italy}

\author{Samuele~Altilia}
\affiliation{Dipartimento di Fisica, Università degli Studi di Milano, Via Celoria 16, Milan, Italy}

\author{Stefano~Olivares}
\affiliation{Dipartimento di Fisica, Università degli Studi di Milano, Via Celoria 16, Milan, Italy}
\affiliation{Istituto Nazionale di Fisica Nucleare, Sezione di Milano, Via Celoria 16, Milan, Italy}

\author{Alessandro~Ferraro}
\affiliation{Dipartimento di Fisica, Università degli Studi di Milano, Via Celoria 16, Milan, Italy}
\affiliation{Istituto Nazionale di Fisica Nucleare, Sezione di Milano, Via Celoria 16, Milan, Italy}

\author{Sebastiano~Corli}
\affiliation{Dipartimento di Fisica, Università degli Studi di Milano, Via Celoria 16, Milan, Italy}

\author{Enrico~Prati}
\affiliation{Dipartimento di Fisica, Università degli Studi di Milano, Via Celoria 16, Milan, Italy}
\affiliation{Istituto Nazionale di Fisica Nucleare, Sezione di Milano, Via Celoria 16, Milan, Italy}
\affiliation{Istituto di Fotonica e Nanotecnologie – CNR, Piazza Leonardo da Vinci 32, I-20133, Milan, Italy}

\author{Simone~Cialdi}
\affiliation{Dipartimento di Fisica, Università degli Studi di Milano, Via Celoria 16, Milan, Italy}
\affiliation{Istituto Nazionale di Fisica Nucleare, Sezione di Milano, Via Celoria 16, Milan, Italy}


\begin{abstract}
An experimental investigation of intracavity dispersion effects in a synchronously pumped optical parametric oscillator (SPOPO) is presented. 
A flexible setup combining spectral and phase shaping of both pump and local oscillator fields with frequency-resolved balanced homodyne detection is employed to examine how intracavity dispersion influences squeezing. 
Different cavity configurations with varying finesse and dispersion conditions are studied, and the squeezing is measured as a function of pump power and local oscillator bandwidth. 
Contrary to expectations based on existing theoretical models, the measured squeezing levels remain essentially unchanged as dispersion varies. 
To account for these observations, a modeling approach is introduced in which intracavity dispersion is described as an effective spectral filtering occurring at the stage of SPOPO supermode generation. 
Within this framework, the filtering is incorporated directly into the interaction Hamiltonian of the nonlinear process. 
This perspective establishes a consistent experimental benchmark for the study of dispersion in SPOPOs and underscores the importance of spectral filtering in the interpretation of multimode squeezing experiments.
\end{abstract}

\maketitle

\section{Introduction}\label{sec:introduction}
The generation and control of squeezed states of light are central to the development of quantum technologies, enabling applications in quantum communication~\cite{Crisafulli2013,Kaiser2016}, boson sampling~\cite{Zhong2021,Madsen2022}, quantum computing with continuous variables~\cite{Asavanant2019,Larsen2019,Walmsley2023}, and quantum-enhanced metrology~\cite{ParisJarda2004,Lawrie2019,Kamble2024,Conlon2024}. 
Multimode squeezed states, in particular, provide access to high-dimensional Hilbert spaces that can be exploited for large-scale entanglement distribution, massively parallel protocols, and the realization of continuous-variable cluster states~\cite{Menicucci2014,Walschaers2019,Walschaers2023}. 
As a result, the ability to generate and manipulate multimode squeezing is a key step towards scalable photonic architectures.

Among the various approaches, the synchronously pumped optical parametric oscillator (SPOPO) is a particularly versatile platform.
Operating in the pulsed regime, it naturally produces a discrete set of squeezed supermodes that can be mapped to quantum frequency combs~\cite{Patera2010,Roslund2014}. 
This property has been exploited in seminal works exploring quantum correlations in the time/frequency domain of SPOPO output~\cite{Jiang2012}, frequency-multiplexed entanglement~\cite{Roslund2014}, and large-scale continuous-variable cluster states~\cite{Asavanant2019,Larsen2019}. 
Such results highlight the SPOPO as a practical source of multimode nonclassical light for applications in quantum networks and computation. 
At the same time, they show that the properties of SPOPO supermodes strongly depend on both the pump spectral profile and the intracavity dynamics.

A critical factor influencing SPOPO operation is intracavity dispersion. 
Ultrashort pulses circulating inside the cavity unavoidably accumulate group-velocity dispersion due to the nonlinear crystal, cavity optics, and air path, thereby reshaping their temporal and spectral profiles. 
Beyond this well-known effect, experiments have shown that cavity dispersion can alter the balance between different frequency components~\cite{Thorpe2008} and thus impact the multimode generation and interactions. 
On the theoretical side, it has been predicted that dispersion modifies the resonance conditions and induces mode-dependent detuning, leading to reduced squeezing, mode coupling, and reduced state purity~\cite{Averchenko2024}. 
These effects are expected to be sensitive to finesse, where for higher finesse pulses circulate longer, and to pump power.
Note that the sensitivity of higher-order modes, with their broader spectral extent, makes them particularly vulnerable to dispersion.

Despite these theoretical predictions, a systematic experimental investigation of dispersion effects in SPOPOs has been lacking. 
Previous experimental works have mostly focused on the demonstration of multimode squeezing and entanglement~\cite{Roslund2014,Asavanant2019,Larsen2019}, or on mode characterization via homodyne tomography and frequency-comb entanglement detection~\cite{Jiang2012,Plick2018}, without directly addressing the interplay between dispersion and multimode structure.

In this work, we report an experimental study of intracavity dispersion effects on multimode squeezing in a SPOPO. 
To the best of our knowledge, no previous experimental work has directly isolated and quantified the impact of intracavity dispersion on squeezing generation in SPOPOs.
We measure the dependence of squeezing on pump power, cavity finesse, and local oscillator bandwidth, comparing compensated and uncompensated dispersion regimes. 
Our results reveal that, contrary to expectations, the measured squeezing is essentially unaffected by dispersion in our experimental conditions.
To interpret our data, we propose to extend the current theoretical model following the approach described in \cite{Christ2014}, i.e., by including a dispersion-induced spectral filtering in the generation process of the supermodes, which narrows the mode spectrum and thereby suppresses the apparent impact of dispersion.
This interpretation reconciles our findings with theory and emphasizes the need to incorporate spectral filtering effects into models of multimode squeezing.

The paper is structured as follows: in Section~\ref{sec:theory} we describe the theoretical model used for the interpretation of our data.
In Section~\ref{sec:setup} we describe our experimental setup and measurement procedures. 
In Section~\ref{sec:measurements} we present and comment the main results, highlighting the unexpected suppression of dispersion effects. 
Section~\ref{sec:conclusions} concludes the paper with concluding remarks.
Finally, we report the formal calculation of the effect of dispersion on cavity modes in Appendix~\ref{appendix_A}.

\section{Theoretical framework}\label{sec:theory}
This work aims to experimentally investigate the effect of intracavity dispersion on squeezing generation in a synchronously pumped optical parametric oscillator.
The theoretical description is based on the model of Averchenko et al.~\cite{Averchenko2024}, which provides detailed simulations of how dispersion modifies the generated squeezing.
Within this framework, the influence of dispersion is quantified by the dimensionless parameter $K = {N_\gamma}/{N_D}$, where $N_\gamma$ is the number of round trips required for the intracavity field to decay to $e^{-1}$ due to cavity losses, scaling with the finesse $\mathcal{F}$ as $N_\gamma = \mathcal{F}/2\pi$, and $N_D$ is the number of round trips for dispersion to broaden the pulse duration by a factor of $\sqrt{2}$, given by $N_D = \tau_s^2 / \mathrm{GDD}$, with $\tau_s$ the temporal duration of the fundamental supermode.
$\mathrm{GDD}$ is the group delay dispersion.
Hence, the parameter $K$ can be written as
\begin{align}\label{eq:K}
    K = \frac{\mathcal{F}}{2\pi}\frac{\mathrm{GDD}}{\tau_s^2} \, .
\end{align}
Measuring squeezing in different regimes of $K$ provides a convenient way to probe the effect of dispersion: for $K \ll 1$, dispersion is expected to be negligible, but becomes relevant already for $K$ near unity.
Experimentally, this can be realized by using SPOPO cavities with different finesse values.

Beyond the cavity finesse, other experimental parameters can be adjusted to probe the effect of intracavity dispersion.
One important parameter is the pump power $P$. 
The dependence on pump power is described in Ref.~\cite{Averchenko2024} via the ratio $\lambda_0/\gamma$, which depends explicitly on $P$ and determines the effective number of contributing modes, being $\gamma$ the loss rate of the cavity.
Another tunable parameter is the spectral bandwidth $\Delta\lambda_\mathrm{LO}$ of the local oscillator (LO) in homodyne, which can be exploited to probe the effect of dispersion, that increases with the mode index \cite{Averchenko2024}.
Indeed, for a symmetric LO phase profile, narrowing the bandwidth increases the projection onto higher-order SPOPO supermodes, thereby enhancing the sensitivity to dispersion.
Measurements can also be performed using an antisymmetric LO phase profile. In fact, a symmetric LO spectrum does not overlap with odd supermodes: introducing a $\pi$ phase jump at the spectral center enables projection onto them.
This selective overlap with odd modes further emphasizes the contribution of higher-order supermodes, making the system particularly sensitive to dispersion effects.
For this reason, measurements that probe these parameters are performed and reported in this work in \textbf{Section}~\ref{sec:measurements}.

\begin{figure}
    \centering
    \includegraphics[width=120mm]{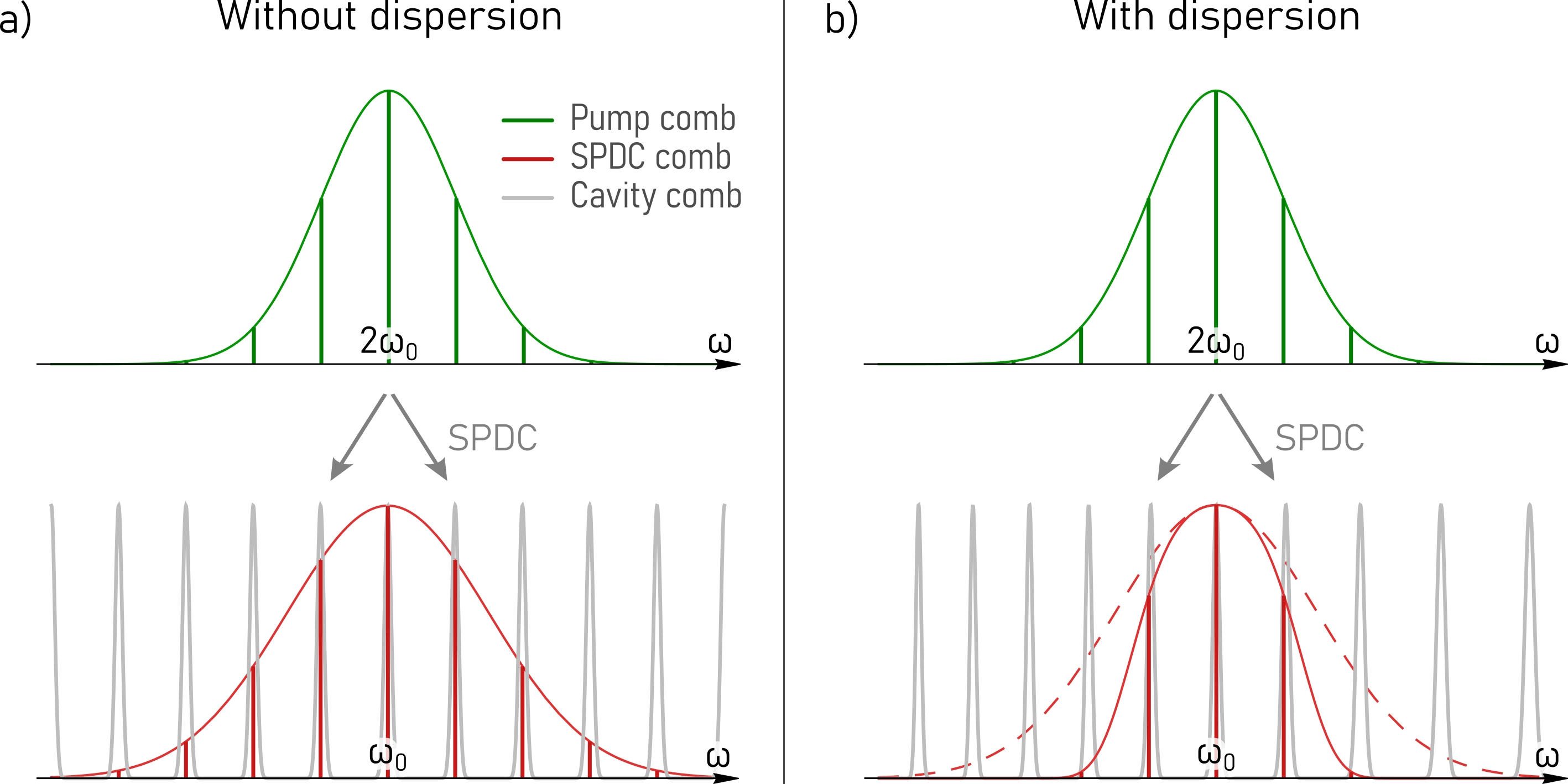}
    \caption{
    Scheme of the combs involved in the SPDC process.
    The pump comb (green), centered around $2\omega_0$, generates the quantum-state comb (red) centered at $\omega_0$.
    The comb of the SPDC photons is then projected onto the cavity comb (gray).
    In the absence of dispersion (a), the quantum-state comb and the cavity comb coincide. 
    In the presence of dispersion (b), a detuning $\delta\omega_j$ arises between the cavity and quantum-state teeth according to Equation~\ref{eq:dispshift}, leading to a spectral cut (notice that the cavity teeth are shifted to right with respect to SPDC teeth).
    The dashed line is the original spectrum for comparison.
    }
    \label{fig:combs}
\end{figure}

On the other hand, our measurements show no observable effect of intracavity dispersion on the measured squeezing levels, in apparent contradiction with the predictions of the theoretical model of Averchenko et al., as we will show in \textbf{Section}~\ref{sec:measurements}.
This discrepancy can be interpreted by noting that intracavity dispersion does not merely affect the subsequent evolution of the generated supermodes, but also plays a crucial role already at the level of their generation, as we will discuss in this section.
To clarify this point, we start by considering the different frequency combs involved in the SPOPO process, shown in \textbf{Figure}~\ref{fig:combs}.
Following the scheme of our experimental setup, which will be discussed in \textbf{Section}~\ref{sec:setup}, we consider a pump radiation (green) at frequency $2\omega_0$, obtained as the second harmonic of an infrared mode-locked source at frequency $\omega_0$.
The latter is also used to actively stabilize the SPOPO, which is therefore resonant with the laser.
The pump comb drives the SPDC process, and energy conservation in SPDC leads to the generation of photon pairs populating a frequency comb centered at $\omega_0$ (red).
Finally, the optical cavity supports its own frequency comb (gray), whose mode frequencies are determined by the cavity round-trip phase condition~\cite{Siegman1986}: it is this cavity comb that is modified by intracavity dispersion.
In the absence of intracavity dispersion (\textbf{Figure}~\ref{fig:combs}a), the cavity comb spacing coincides with the quantum-state comb generated by SPDC, so that all generated frequency components are resonant with the cavity.
In contrast, when intracavity dispersion is present (\textbf{Figure}~\ref{fig:combs}b), the cavity resonance frequencies are shifted with respect to the SPDC comb.
This mismatch introduces a frequency-dependent detuning $\delta\omega_j$ between the $j$-th cavity mode and the corresponding SPDC frequency component~\cite{Thorpe2008}, leading to an effective spectral filtering already at the generation stage.
Following the result derived in Appendix~\ref{appendix_A}, the detuning $\delta\omega_j$ is given by
\begin{align}\label{eq:dispshift}
    \delta\omega_j = \frac{1}{2}\,\mathrm{GDD}\,\mathrm{FSR}\,\omega_j^2 \, ,
\end{align}
where $\omega_j$ denotes the frequency of the $j$-th tooth with respect to the central frequency $\omega_0$, and $\mathrm{FSR}$ is the cavity free spectral range.
As a consequence, the probability amplitude for generating photons in a given cavity mode is reduced when the corresponding SPDC frequency component falls outside the cavity linewidth.

\begin{figure}
    \centering
    \includegraphics[width=60mm]{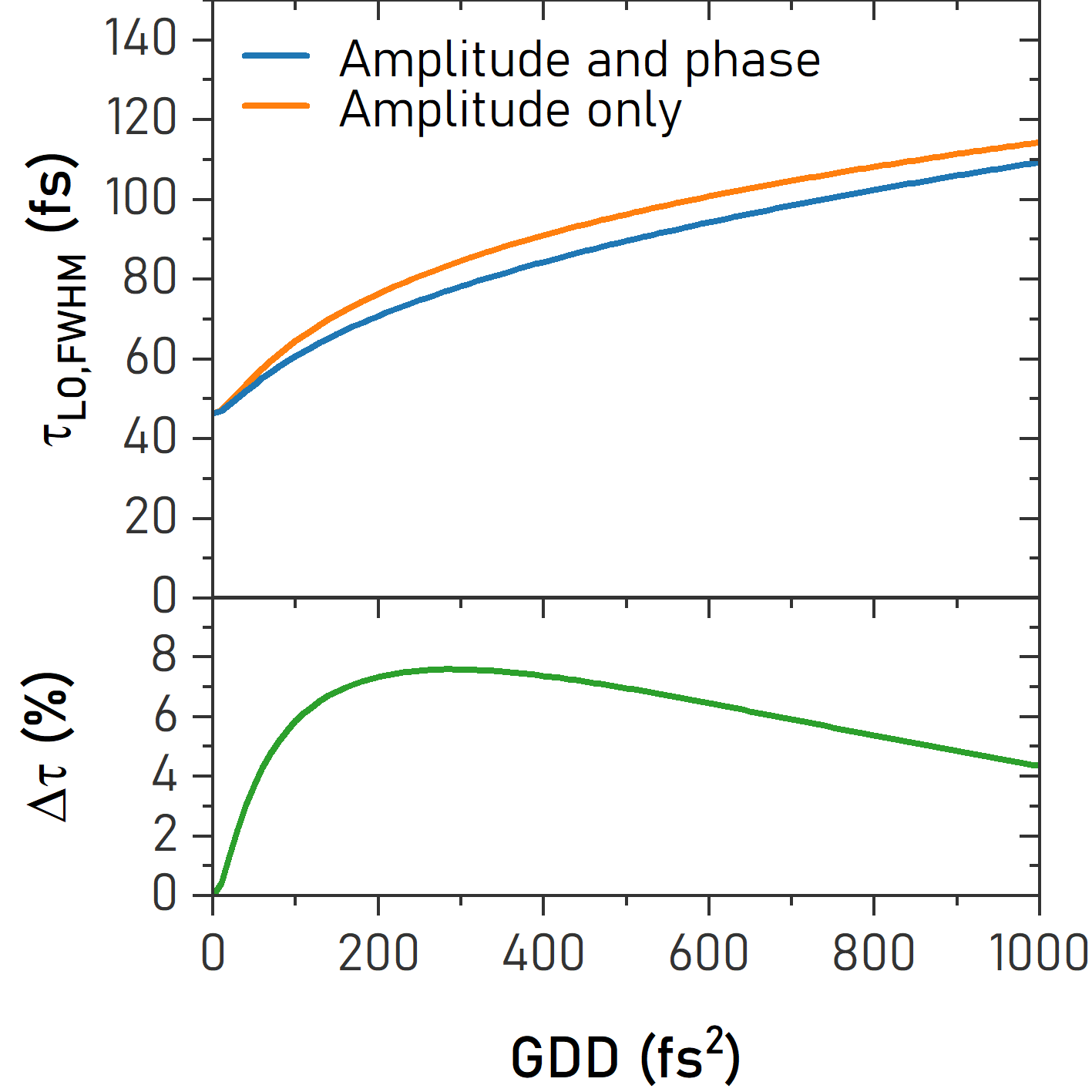}
    \caption{
    Assessment of the impact of the dispersive phase contribution $\Phi(\omega)$ under the experimental parameters of this work.
    Top: FWHM in the temporal domain obtained from the Fourier transform of the full cavity response $f(\omega)$ (blue) and of the amplitude term $A(\omega)$ alone (orange), as a function of the GDD.
    Bottom: relative difference in the temporal width between the two cases (green).
    At a GDD of \SI{900}{\femto\second\squared}, neglecting the phase term introduces an error below \SI{5}{\percent}, and in all cases the deviation remains below \SI{8}{\percent}, quantitatively validating the amplitude-only approximation.
    }
    \label{fig:chirpEA}
\end{figure}

More explicitly, the complex field modulation introduced by an optical cavity (see, for example, Ref.~\cite{Svelto2010} p.~143) can be written, after straightforward algebra, in polar form as
\begin{align}\label{eq:cavmodulation}
    f(\omega) =
    A(\omega)\,\exp\!\left[i\,\Phi(\omega)\right]
    \propto \frac{1}{\left|1-\sqrt{R}\,e^{i \phi(\omega)}\right|}
    \exp\!\left[i\,\arctan\!\left(\frac{\sqrt{R}\sin\phi(\omega)}{1-\sqrt{R}\cos\phi(\omega)}\right)\right] \, ,
\end{align}
where $\omega$ denotes the frequency shift with respect to the central frequency $\omega_0$, $R=\prod_i R_i$ is the total cavity reflectivity.
In case of dispersive cavity, we have a dispersive phase $\phi(\omega)=\tfrac{1}{2}\,\mathrm{GDD}\,\omega^2$ associated with Equation~\ref{eq:dispshift} (a phase shift of $2\pi$ corresponds to a frequency shift of one free spectral range).
Under the parameters of our experiment, the phase contribution $\Phi(\omega)$ is sufficiently small that it can be safely neglected without affecting the relevant physical conclusions, allowing us to consider only the amplitude modulation $A(\omega)$ introduced by dispersion.
This is verified by comparing the temporal duration of the Fourier transform of the full cavity response $f(\omega)$ and the amplitude term $A(\omega)$ alone.
As shown in \textbf{Figure}~\ref{fig:chirpEA}, the upper panel reports the full width at half maximum in the temporal domain as a function of the group-delay dispersion of full function $f(\omega)$ (blue) and of the amplitude term $A(\omega)$ only (orange).
The lower panel shows the relative difference in the temporal width of the Fourier transform of 
$f(\omega)$ between the two cases (green): for a GDD of \SI{900}{\femto\second\squared}, the error on the temporal duration introduced by neglecting the phase term remains below \SI{5}{\percent}, and always stays below the \SI{8}{\percent}.
This quantitatively justifies the use of the amplitude-only approximation under our experimental conditions.

For this reason, when considering the power spectral density of the individual comb teeth, defined as $F(\omega)=|f(\omega)|^2$, one can use the approximate expression
\begin{align}
    F(\omega) \propto \frac{1}{\left|1 - \sqrt{R} e^{i\,\phi\left(\omega\right)}\right|^2} = \frac{1}{1 + R - 2\sqrt{R}\cos{\phi\!\left(\omega\right)}} \, .
\end{align}
Since the cavity finesse is high, each comb tooth is spectrally narrow, and we operate in the vicinity of the cavity resonances.
As a result, the phase $\phi(\omega)$ varies only weakly over the relevant spectral range, i.e.\ $\Delta\phi \ll 2\pi$, allowing us to expand the cosine function to second order.
After straightforward algebra, this yields a Lorentzian modulation
\begin{align}
    F(\omega) \propto \frac{1}{1 + \frac{\pi^2}{\mathcal{F}^2}\phi^2\!\left(\omega\right)} \, ,
\end{align}
where $\mathcal{F} = \frac{\pi R^{1/4}}{1-\sqrt{R}}$ is the cavity finesse.
By explicitly inserting the dispersive phase $\phi(\omega)=\tfrac{1}{2}\mathrm{GDD}\,\omega^2$, this expression becomes
\begin{align}\label{eq:intensitygdd}
    F(\omega) \propto \frac{1}{1 + \frac{\pi^2}{4\mathcal{F}^2}\mathrm{GDD}^2\,\omega^4} \, ,
\end{align}
corresponding to a full width at half maximum of $\Delta\omega_\mathrm{FWHM}=\sqrt{\frac{8\,\mathcal{F}}{\pi\,\mathrm{GDD}}}$, which, when expressed in wavelength units, reads $\Delta\lambda_\mathrm{FWHM} = \frac{\lambda_0^2}{c}\sqrt{\frac{2\,\mathcal{F}}{\pi^3\,\mathrm{GDD}}}$, being $\lambda_0$ the central wavelength.
This shows explicitly that intracavity group-delay dispersion acts as a spectral filter on the cavity-supported modes.

At this point, spectral filtering can be included in the model of SPDC following the same approach proposed in \cite{Christ2014}.
In particular, they propose to apply the filtering function directly to the Joint Spectral Amplitude (JSA) of the original states (i.e., the states generated with no dispersion), and then apply a Singular Value Decomposition (SVD) to find the new filtered supermodes of the SPDC.
As demonstrated in \cite{Christ2014}, this procedure yields a quantitatively accurate description of spectral filtering in SPDC.

For this reason, we incorporate the filtering of Equation~\ref{eq:intensitygdd} directly into the interaction Hamiltonian.
The interaction among SPOPO modes is described by~\cite{Roslund2014}
\begin{align}\label{eq:hamiltonian}
    \hat{H} = i\hbar\,g\sum_{m,n} L_{m,n}\,\hat{a}_m^{\dagger}\hat{a}_n^{\dagger} + \mathrm{h.c.} \, ,
\end{align}
where $\hat{a}_j^\dagger$ is the creation operator for a photon in the $j$-th cavity mode at frequency $\omega_j$ (measured with respect to $\omega_0$), and $g$ is the interaction constant.
The coupling matrix
\begin{align}
    L_{m,n} = p_{m,n}\,A^{(p)}_{m+n}
\end{align}
combines the nonlinear crystal phase-matching function $p_{m,n}$~\cite{Grice2001,Mosley2008} with the pump spectral amplitude $A^{(p)}_{m+n}$ at frequency $\omega_m+\omega_n$~\cite{Patera2010}.
In the presence of intracavity dispersion, the generation probability for each photon pair must be projected onto the cavity mode profiles.
This is accounted for by introducing a generalized coupling matrix
\begin{align}\label{eq:Lmatrixfil}
    L'_{m,n} = p_{m,n}\,A^{(p)}_{m+n}\,d_{m,n} \, ,
\end{align}
with $d_{m,n} = \sqrt{F_m}\,\sqrt{F_n}$.
In this framework, intracavity dispersion constrains the generation of supermodes by suppressing frequency components that are not resonant with the cavity comb.

\begin{figure}
    \centering
    \includegraphics[width=60mm]{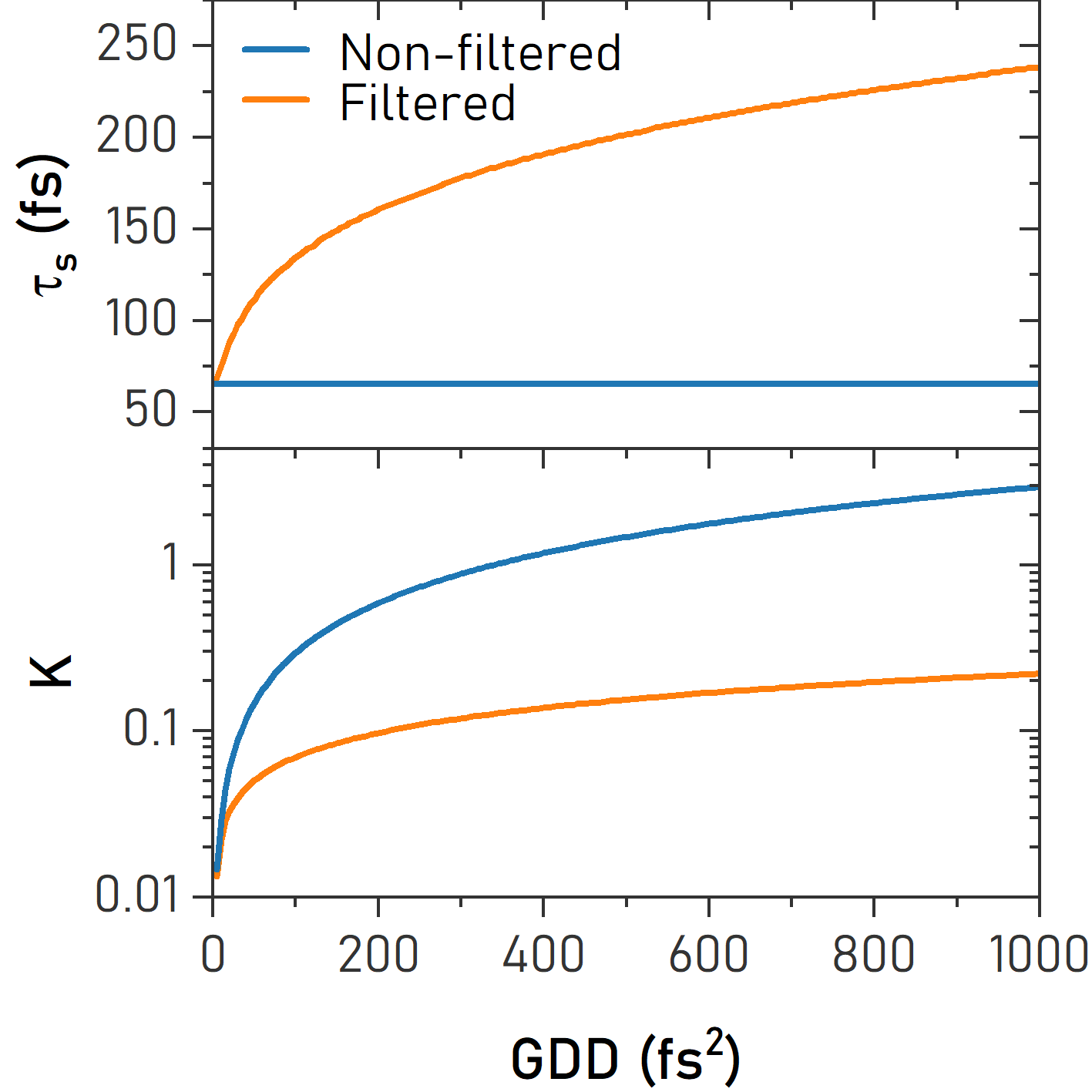}
    \caption{
    Top: temporal duration $\tau_s$ of the fundamental supermode as a function of intracavity dispersion $\mathrm{GDD}$, calculated from the unfiltered matrix $L_{m,n}$ (blue) and from the filtered matrix $L'_{m,n}$ (orange). 
    Bottom: corresponding parameter $K$ for the two models. 
    The cavity finesse is $\mathcal{F} = 75$.
    }
    \label{fig:vsGDD}
\end{figure}

The introduction of spectral filtering modifies the temporal structure of the SPOPO supermodes: applying a singular value decomposition (SVD) to the filtered coupling matrix $L'_{m,n}$ yields a new set of supermodes $\hat{S}_k'^\dagger$, each associated with an eigenvalue $\Delta'_k$~\cite{Braunstein2005,Patera2010,Fabre2020}, whose temporal duration is stretched compared to the original modes $\hat{S}_k^\dagger$ obtained from $L_{m,n}$ without filtering. 
Consequently, the parameter $K$ of Equation~\ref{eq:K} no longer scales linearly with $\mathrm{GDD}$, but exhibits a slower growth, as $\tau_s^2$ in the denominator increases with filtering. 
This effect is illustrated in \textbf{Figure}~\ref{fig:vsGDD}, where the top panel shows the temporal duration of the fundamental supermode as a function of intracavity dispersion, while the bottom panel reports the corresponding parameter $K$. 
In the filtered model, the effective impact of dispersion is significantly reduced. 
For example, under our experimental conditions with $\mathrm{GDD} = \SI{900}{\femto\second\squared}$, the effect of dispersion calculated including spectral filtering is equivalent to only about \SI{70}{\femto\second\squared} in the unfiltered model, i.e., negligible in practice. 
This provides a natural explanation for the absence of observable differences between compensated and uncompensated dispersion in our squeezing measurements, reported in \textbf{Section}~\ref{sec:measurements}.
Notice that intracavity dispersion does not have a negligible effect in absolute terms, rather, its impact is negligible under the conditions of our configuration.
In different regimes characterized by significantly larger dispersion values (well beyond those typically encountered in standard experimental implementations), its influence would become observable. 
Indeed, as shown in the lower panel of \textbf{Figure}~\ref{fig:vsGDD}, the orange curve, although lying below the blue one, exhibits a clear increasing trend with GDD. 
This behavior indicates that for sufficiently large dispersion the deviation would become appreciable, and measurable effects on the squeezing level would eventually emerge.

It is worth noting that the local oscillator (LO) undergoes the same intracavity dispersion as the generated quantum state, since both originate from the same optical cavity. 
The same argument discussed above (see Equation~\ref{eq:cavmodulation}) therefore applies to the LO: dispersion acts as a spectral filter on it. 
However, its bandwidth is initially limited to values below \SI{3}{\nano\meter}, so that under our experimental conditions the resulting spectral filtering remains negligible. 
If the effects described in~\cite{Averchenko2024} were present, they would be revealed by the LO, since the homodyne projection involves multiple supermodes experiencing different dispersive contributions.

In homodyne detection, the measured squeezing corresponds not to a single SPOPO supermode, but to a weighted combination of all contributing modes. 
Formally, the detected quadrature variance can be expressed as~\cite{Patera2010}:
\begin{align}\label{eq:sqtot}
    \sigma^2_{\mathrm{s}} = 
    \sum_{k=0}^{\infty} \left|M_k\right|^2 \,\sigma^2_{\mathrm{sq},k} \, ,
\end{align}
where $M_k$ quantifies the overlap between the local oscillator and the $k$-th supermode, and $\sigma^2_{\mathrm{sq},k}$ is its quadrature  (with shot-noise variance set to $1$).
Here, we considered a LO of the form either \\$g_s\!\left(\omega\right) \propto \exp\!\left[-4 \ln 2 \, \omega^2 / \mathrm{FWHM}^2 \right]$, or $g_a\!\left(\omega\right) \propto \exp\!\left[-4 \ln 2 \, \omega^2 / \mathrm{FWHM}^2 \right]\operatorname{sgn}\!\left(\omega\right)$, for symmetrical and anti-symmetrical LO, respectively. 
$\mathrm{FWHM}$ denotes the full width at half maximum of the LO spectrum, while $\operatorname{sgn}\!\left(\omega\right)={\omega}/{\left|\omega\right|}$.
For numerical stability, it is convenient to truncate the sum in Equation~\ref{eq:sqtot} at a mode index $k_0$, beyond which the quadrature variance is indistinguishable from vacuum.
This yields
\begin{align}\label{eq:sqtottruncated}
    \sigma^2_{\mathrm{s}} = 
    \sum_{k=0}^{k_0} \left|M_k\right|^2 \,\sigma^2_{\mathrm{sq},k}
    +
    \left(1 - \sum_{k=0}^{k_0} \left|M_k\right|^2\right) \, ,
\end{align}
The first term of Equation~\ref{eq:sqtottruncated} accounts for the contribution of the first $k_0$-th modes projected onto the LO, while the second term represents the residual fraction $k>k_0$, which contributes as vacuum noise with unit variance. 
In all simulations we use $k_0=150$.

\begin{figure}
    \centering
    \includegraphics[width=100mm]{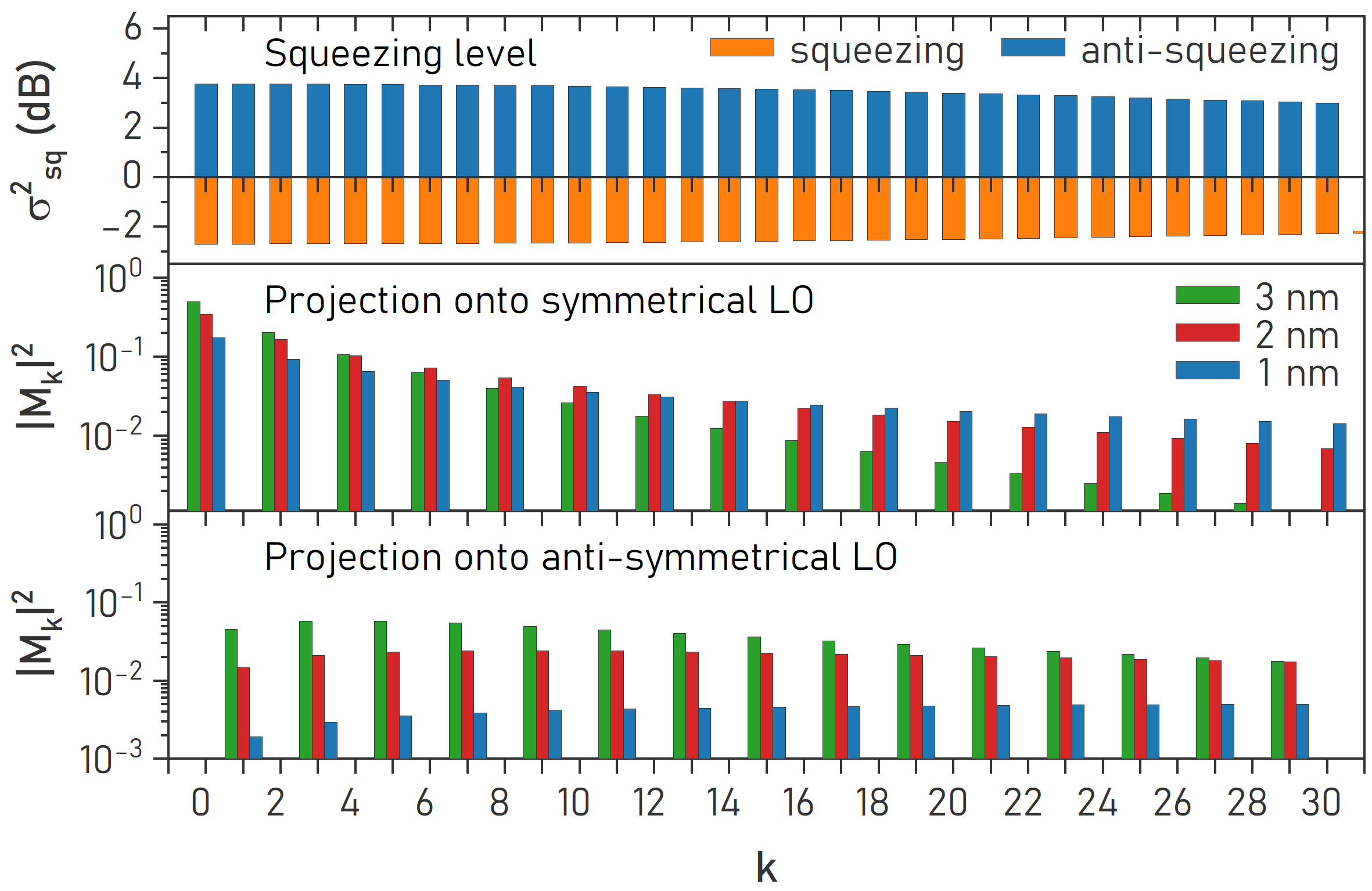}
    \caption{
    Simulations for no intracavity dispersion.
    Top: calculated squeezing and anti-squeezing levels $\sigma_\mathrm{sq}^2$ at SPOPO output for the first 30 supermodes $k$ at $P=0.3$. 
    Middle: projections $\left|M_k\right|^2$ onto symmetric LOs with FWHM bandwidths of \SI{1}{\nano\meter}, \SI{2}{\nano\meter}, and \SI{3}{\nano\meter}. 
    Bottom: projections for antisymmetric LOs with the same bandwidths. 
    Simulations assume $R_\mathrm{oc}=0.95$ and $\mathrm{GDD}=\SI{0}{\femto\second\squared}$.
    Note that modes up to high orders contribute to the total squeezing level at SPOPO output. 
    It is also clear that the projection onto the LO extends to higher order modes by reducing LO spectral width, or by using an antisymmetrical LO.
    }
    \label{fig:proj_LO}
\end{figure}

\begin{figure}
    \centering
    \includegraphics[width=100mm]{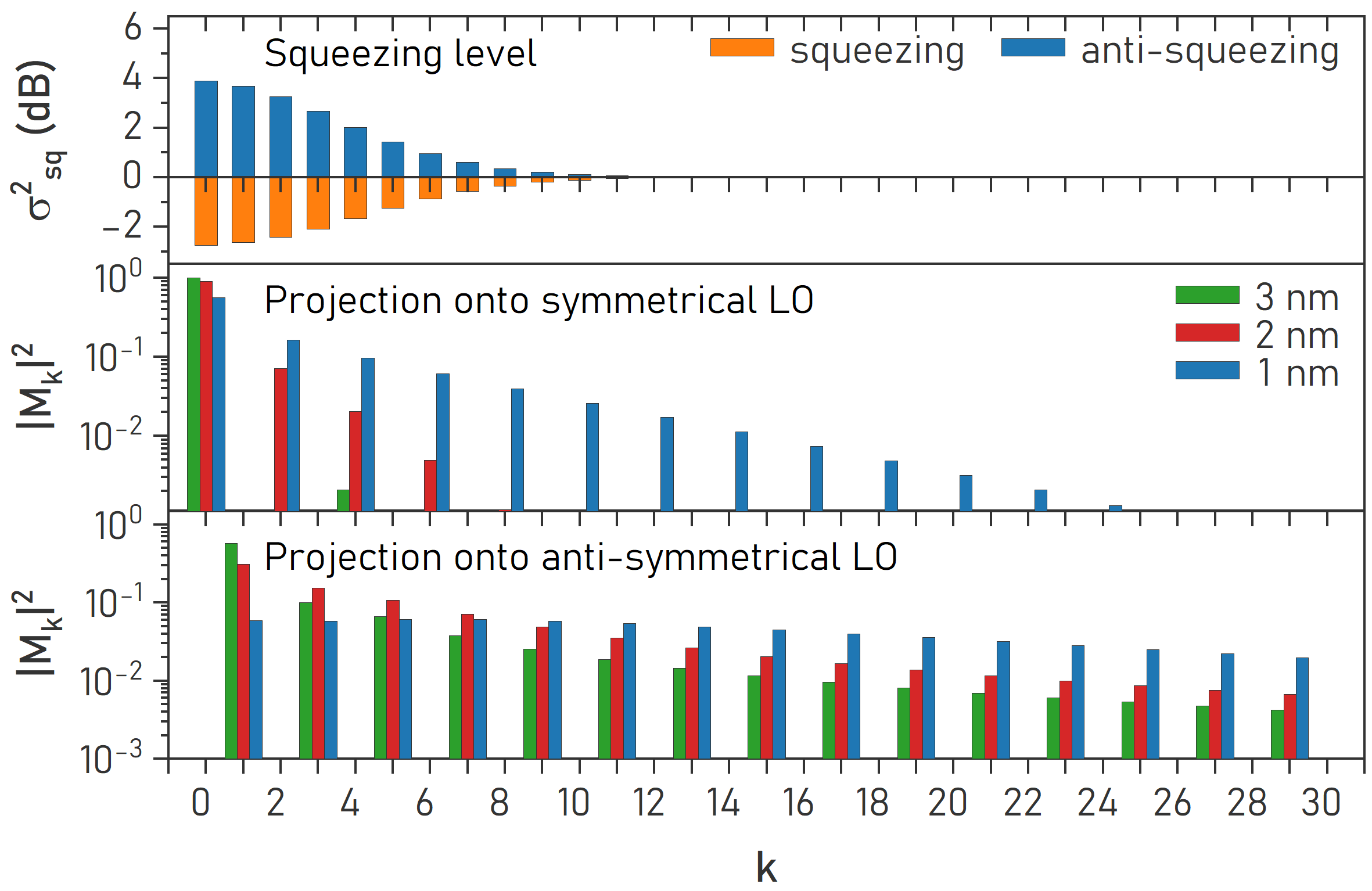}
    \caption{
    Simulations for \SI{900}{\femto\second\squared} of intracavity dispersion.
    Top: calculated squeezing and anti-squeezing levels $\sigma_\mathrm{sq}^2$ at SPOPO output for the first 30 supermodes $k$ at $P=0.3$. 
    Middle: projections $\left|M_k\right|^2$ onto symmetric LOs with FWHM bandwidths of \SI{1}{\nano\meter}, \SI{2}{\nano\meter}, and \SI{3}{\nano\meter}. 
    Bottom: projections for antisymmetric LOs with the same bandwidths. 
    Simulations assume $R_\mathrm{oc}=0.95$ and $\mathrm{GDD}=\SI{900}{\femto\second\squared}$.
    Note that only the first few modes contribute to the total squeezing level at SPOPO output, as the variance of higher order modes approaches the shot noise level.
    Also in this case, the projection onto the LO extends to higher order modes by reducing LO spectral width, or by using an antisymmetrical LO.
    }
    \label{fig:proj_LO-GDD900}
\end{figure}

\textbf{Figure}~\ref{fig:proj_LO} and \textbf{Figure} \ref{fig:proj_LO-GDD900} illustrate the simulation of the squeezing level of the first 30 modes, and their projection onto a symmetrical and an antisymmetrical LO, without intracavity dispersion and with \SI{900}{\femto\second\squared} of intracavity dispersion, respectively.
These simulations were performed following the procedure described in \cite{Fabre2020}: after performing a singular-value decomposition (SVD), we computed the squeezing level of each mode $k$ at SPOPO output at $P=0.3$ (top), and the corresponding projections $\left|M_k\right|^2$ onto symmetric and antisymmetric LOs $g_{s,a}\!\left(\omega\right)$, respectively, with FWHM spectral bandwidths of \SI{1}{\nano\meter}, \SI{2}{\nano\meter}, and \SI{3}{\nano\meter} (center and bottom). 
For all simulations presented in this paper, the frequency $\omega$ was sampled at 1600 points spanning \SI{-160}{\radian\per\pico\second} to \SI{160}{\radian\per\pico\second} for the SVD computation. 
In the case of no intracavity dispersion (\textbf{Figure}~\ref{fig:proj_LO}), modes up to high orders contribute to the total squeezing level at SPOPO output. 
It is also clear that the projection onto the LO extends to higher order modes by reducing LO spectral width, or by using an antisymmetrical LO.
On the other hand, in the case of \SI{900}{\femto\second\squared} intracavity dispersion (\textbf{Figure}~\ref{fig:proj_LO-GDD900}) only the first few modes contribute to the total squeezing level at SPOPO output, as the variance of higher order modes approaches the shot noise level.
Also in this case, the projection onto the LO extends to higher order modes by reducing LO spectral width, or by using an antisymmetrical LO.
These results provide a clear rationale for scanning the LO bandwidth and phase symmetry: reducing $\Delta\lambda_\mathrm{LO}$ or using an antisymmetric LO increases overlap with modes that are theoretically most sensitive to intracavity dispersion. 

Notice that the simulations also provide an intuitive interpretation of the weak impact of intracavity dispersion observed in our configuration. 
In the absence of GDD, significant squeezing is distributed over several supermodes, and the projection of the local oscillator onto these modes is likewise substantial over a broad range of mode orders. 
As a result, the measured squeezing arises from the contribution of multiple modes.
On the other hand, when GDD is introduced, the squeezing level decreases rapidly with increasing supermode index, but, at the same time, the projection of the LO onto higher-order modes decreases.
Consequently, the reduction of squeezing in higher-order modes is accompanied by a corresponding reduction of their weight in the homodyne measurement.
These two effects compensate each other, leading to nearly identical total measured squeezing levels with and without dispersion.

\section{Experimental setup}\label{sec:setup}
\begin{figure}
    \centering
    \includegraphics[width=120mm]{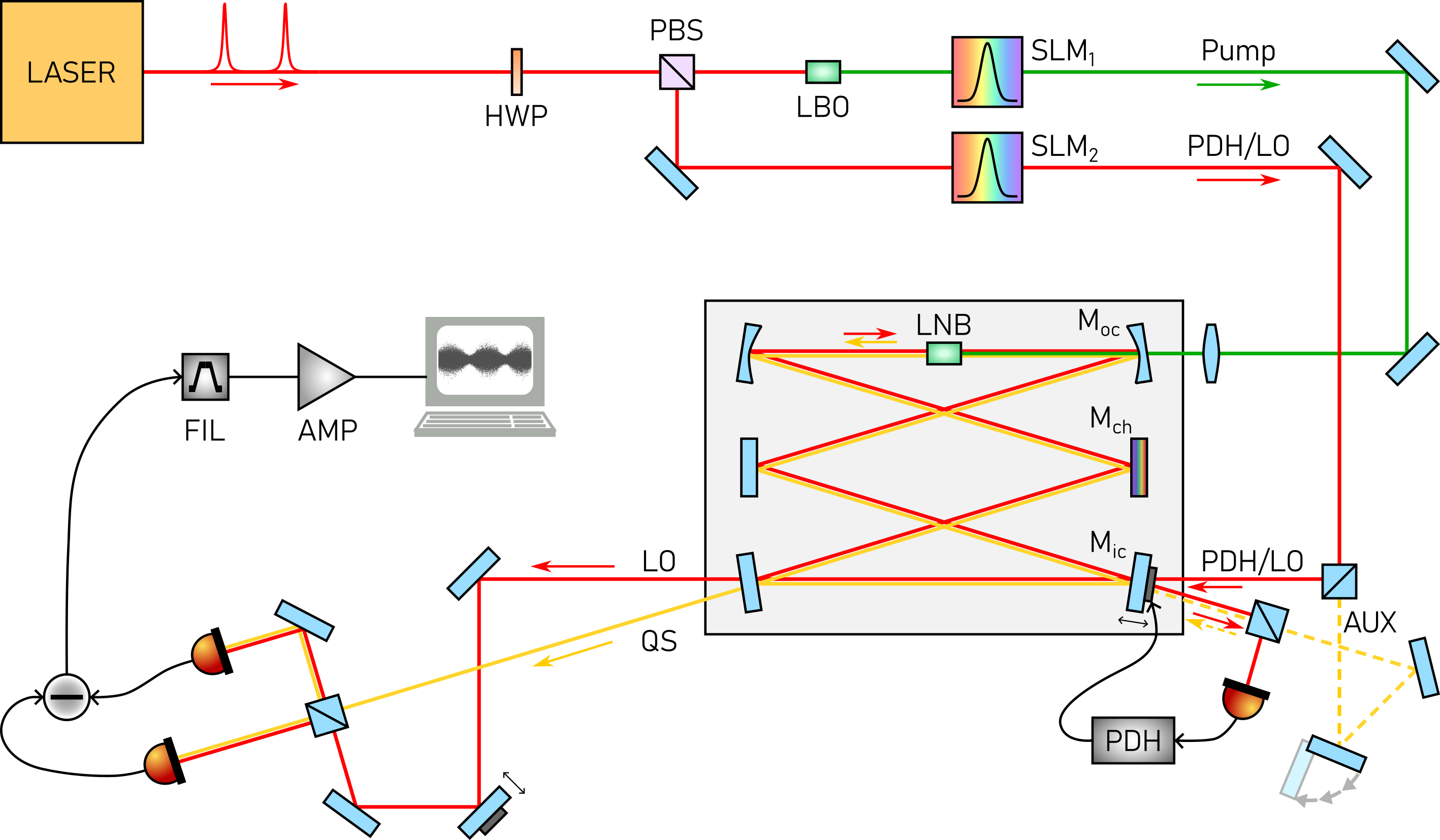}
    \caption{
    Scheme of the experimental setup.
    The pump and the local oscillator (LO) are derived from a \SI{93}{\mega\hertz} mode-locked laser and are independently shaped in spectral amplitude and phase using two spatial light modulators (SLMs).
    The SPOPO is frequency-stabilized to the laser repetition rate, and both the generated quantum state and the LO originate from the same cavity mode.
    A balanced homodyne detection stage is used to measure the frequency-resolved squeezing level.
    An auxiliary beam (AUX), derived from the PDH/LO beam, is overlapped with the intracavity SPDC field and serves as a weak external seed on demand.
    It is employed to measure both the parametric gain of the SPOPO and the visibility of the homodyne detector after transmission through the cavity.
    HWP: half-waveplate, PBS: polarizing beam splitter, LBO: lithium-borate crystal, SLM$_\mathrm{i}$: spatial light modulators, PDH: Pound-Drever-Hall electronics, M$_\mathrm{ic}$ and M$_\mathrm{oc}$: input and output couplers, M$_\mathrm{ch}$ chirped mirror, LNB: lithium niobate crystal, LO: local oscillator, QS: quantum state, FIL: bandpass filter, AMP: differential amplifier.
    }
    \label{fig:setup}
\end{figure}

In this section, we describe the experimental setup used for the generation and detection of pulsed squeezed quantum states for the study of intracavity dispersion effects in the SPOPO.
The setup, illustrated in \textbf{Figure}~\ref{fig:setup}, is described in~\cite{Suerra2026}, and consists of three main parts: a laser source providing both pump and local oscillator radiation, a SPOPO, and a balanced homodyne detection system.
The source is a commercial mode-locked Ytterbium-doped fiber laser with a repetition rate of \SI{93}{\mega\hertz}, a central wavelength of \SI{1035}{\nano\meter}, and a spectral width of nearly \SI{10}{\nano\meter}.
Its output is split into two arms: one is frequency-doubled in a \SI{5.0}{\milli\meter} LBO crystal to generate \SI{517.5}{\nano\meter} pump pulses for the SPOPO; the other serves both for frequency locking of the SPOPO and, after transmission through the cavity, as the LO for homodyne detection.
Both IR and green beams are independently shaped in amplitude and phase using two spatial light modulators (SLMs) in a similar manner of that described in~\cite{Monmayrant2010}.
The pump spectrum is tailored to a Gaussian profile with flat phase and FWHM of \SI{1.0}{\nano\meter}, yielding near-transform-limited pulses of \SI{400}{\femto\second}-FWHM duration.
The IR beam can be spectrally filtered from \SI{0.5}{\nano\meter} up to \SI{3.0}{\nano\meter} FWHM, maintaining near-transform-limited pulses.
These spectral constraints arise from spatial inhomogeneities in the laser output~\cite{Suerra2025}, which require selecting a narrower spectral region via the SLMs.

The SPOPO cavity consists of six mirrors in a near-confocal geometry, with four flat and two curved mirrors. 
The input coupler has a reflectivity of \SI{99}{\percent}, while the output coupler is chosen as either \SI{81}{\percent} or \SI{95}{\percent} to vary the cavity finesse and escape efficiency. 
All other mirrors have high reflectivity. 
A \SI{3.0}{\milli\meter}-long lithium niobate nonlinear crystal, used for type-I parametric down-conversion, is positioned at the cavity waist between the curved mirrors, where the mode radius is approximately \SI{80}{\micro\meter}. 
The pump beam is focused to a waist of \SI{50}{\micro\meter} with adjustable power up to \SI{80}{\milli\watt}.
The SPOPO is frequency-locked to the laser using the IR beam from SLM$_2$ and a standard Pound–Drever–Hall technique~\cite{Drever1983}.
The counter-propagating auxiliary beam at the same wavelength as the quantum state naturally separates the locking beam from the squeezed output and simultaneously serves as the LO.

Intracavity dispersion is controlled by replacing one of the flat HR mirrors with a chirped mirror.
In this work, the dispersion introduced by the \SI{3.0}{\milli\meter}-long lithium niobate crystal ($\sim\SI{850}{\femto\second\squared}$) and the air path ($\sim\SI{50}{\femto\second\squared}$) was either fully compensated with a chirped mirror providing \SI{-900}{\femto\second\squared} or left uncompensated (\SI{0}{\femto\second\squared}), yielding a net GDD of zero or \SI{900}{\femto\second\squared}, respectively.

Quantum states are detected using a standard frequency-resolved homodyne detection scheme~\cite{Bachor2019}.  
The auxiliary IR beam from SLM$_2$ serves as the LO, ensuring near-ideal spatial mode overlap with the quantum state, as both originate from the same cavity mode. 
In addition to the symmetric Gaussian profile, the LO can also be shaped into an anti-symmetric mode by introducing a $\pi$ phase jump at the spectral center via the SLM$_2$.
This operation allows projection onto the odd supermodes of the SPOPO, in contrast to the even-mode projection achieved with a symmetric LO.  
Signal and LO are mixed on a 50:50 beam splitter, and the outputs are detected by a balanced photodiode pair.  
The difference signal is amplified and filtered with a bandpass filter between \SI{400}{\kilo\hertz} and \SI{600}{\kilo\hertz}, within the SPOPO bandwidth of either \SI{3.9}{\mega\hertz} or \SI{1.2}{\mega\hertz} (depending on the finesse) and above mechanical noise.
The shared origin of the LO and quantum state ensures stable overlap, yielding high homodyne visibility, $\geq \SI{94}{\percent}$.
A delay line synchronizes the LO and signal pulses temporally.
Another auxiliary beam (AUX), derived from the PDH/LO beam, is overlapped with the intracavity SPDC field and serves as a weak external seed on demand.
It is employed to measure both the parametric gain of the SPOPO and the visibility of the homodyne detector after transmission through the cavity.

To retrieve the intrinsic squeezing produced by the SPOPO, the measured variance $\sigma^2_\mathrm{hom}$ must be corrected for the finite detection efficiency $\eta_\mathrm{hom}$~\cite{Leonhardt1995,DARIANO2003,Olivares21} (we set to 1 the shot-noise variance):
\begin{align}\label{HD:var}
    \sigma^2_\mathrm{hom}(\theta) = 
    1 - \eta_\mathrm{hom}\left(1 - \sigma^2_\mathrm{s}(\theta)\right) 
    \, ,
\end{align}
where $\sigma^2_\mathrm{s}(\theta)$ denotes the actual variance of the SPOPO output state generated inside the cavity for a given quadrature $X(\theta)$, projected onto our LO. 
The overall detection efficiency results from the product of several factors~\cite{Roman-Rodriguez2024}:
\begin{align}
    \eta_\mathrm{hom} = \eta_\mathrm{PD} \, \eta_\mathrm{opt} \, \eta_\mathrm{vis} \, \eta_\mathrm{bkg} \, ,
\end{align}
Here, $\eta_\mathrm{PD} = \SI{87}{\percent}$ is the quantum efficiency of the photodiodes, $\eta_\mathrm{opt} \sim \SI{99}{\percent}$ accounting for optical transmission losses and non-ideal beam splitting, and $\eta_\mathrm{vis} = \mathrm{vis}^2$, where $\mathrm{vis} \geq \SI{94}{\percent}$ is the measured visibility. 
The factor $\eta_\mathrm{bkg}$ accounts for the residual contribution of electronic noise and imperfect common-mode rejection in the balanced detector. 
In practice, $\eta_\mathrm{bkg}$ is determined by comparing the measured electronic-noise floor, the difference-current signal (i.e., the homodyne output), and the single-input signal (from one photodiode). 
This standard procedure in balanced homodyne detection quantifies the residual background as an effective loss channel, yielding $\eta_\mathrm{bkg} \approx 0.96$ in our detection band. 
This corresponds to a common-mode rejection ratio (CMRR) of nearly \SI{54}{\decibel}, achieved through careful optimization of the differential amplifier circuitry.
Such a high level of rejection is crucial for suppressing technical noise originating from the commercial laser source.

\begin{figure}
    \centering
    \includegraphics[width=60mm]{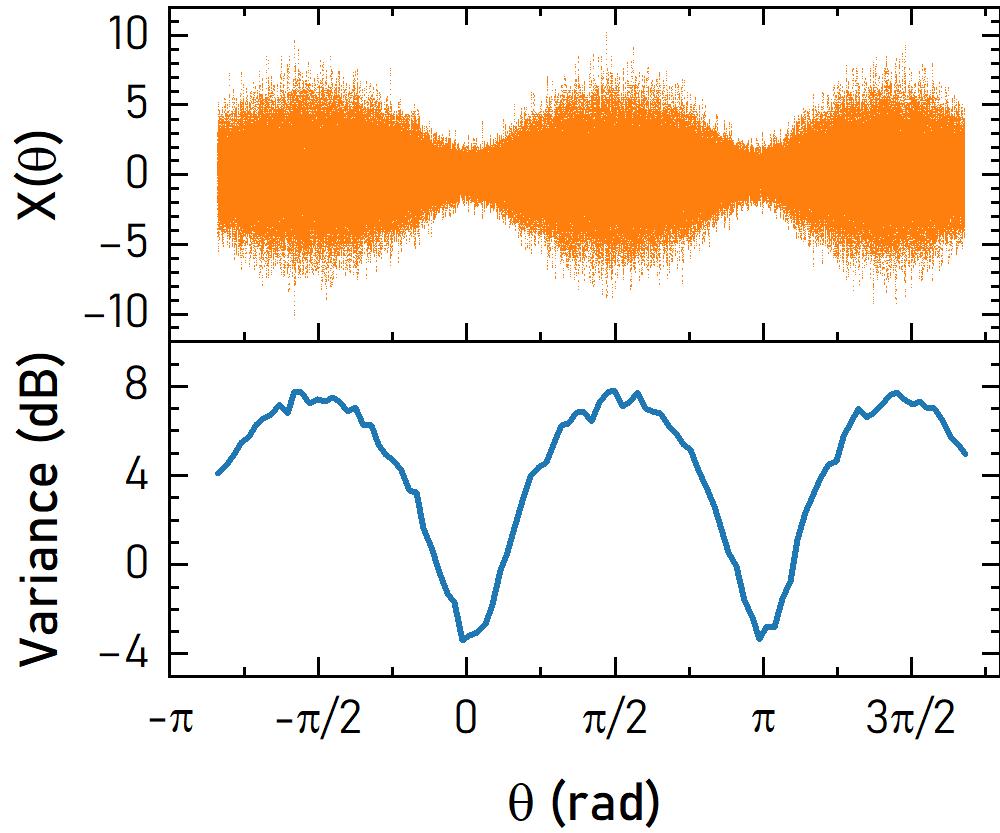}
    \caption{
    Example of homodyne trace (orange, above) and calculated squeezing level (blue, below), measured with our setup.
    Here, the pump power is $P\approx0.3$, $R_\mathrm{oc}=0.81$, $\mathrm{GDD}=\SI{0}{\femto\second\squared}$, and the LO bandwidth is \SI{3}{\nano\meter}-FWHM.
    }
    \label{fig:homodynetrace}
\end{figure}

An example of homodyne trace measured with our setup is shown in \textbf{Figure} \ref{fig:homodynetrace}.

\section{Measurements}\label{sec:measurements}
As anticipated in \textbf{Section}~\ref{sec:theory}, the effect of intracavity dispersion has been probed by measuring squeezing and anti-squeezing as functions of the pump power and the spectral bandwidth of the local oscillator.  
Measurements were performed for two output-coupler reflectivities, $R_{\mathrm{oc}}=0.81$ and $R_{\mathrm{oc}}=0.95$, which correspond to cavity finesse experimental values $\mathcal{F}=24$ and $\mathcal{F}=75$, respectively.
For each reflectivity, we acquired data in two dispersion regimes, for a direct comparison: dispersion-compensated (net $\mathrm{GDD}=\SI{0}{\femto\second\squared}$) and uncompensated (net $\mathrm{GDD}=\SI{900}{\femto\second\squared}$).
Quantitatively, using our experimental parameters and the interaction matrix $L_{m,n}$, one would expect $K=0.83$ for $R_\mathrm{oc} = 0.81$ and $\mathrm{GDD} = \SI{900}{\femto\second\squared}$, and $K = 2.6$ for $R_\mathrm{oc} = 0.95$ and $\mathrm{GDD} = \SI{900}{\femto\second\squared}$.
In each case, we should be in a regime in which dispersion is expected to produce measurable effects following the model of~\cite{Averchenko2024}.

\begin{figure}
    \centering
    \includegraphics[width=120mm]{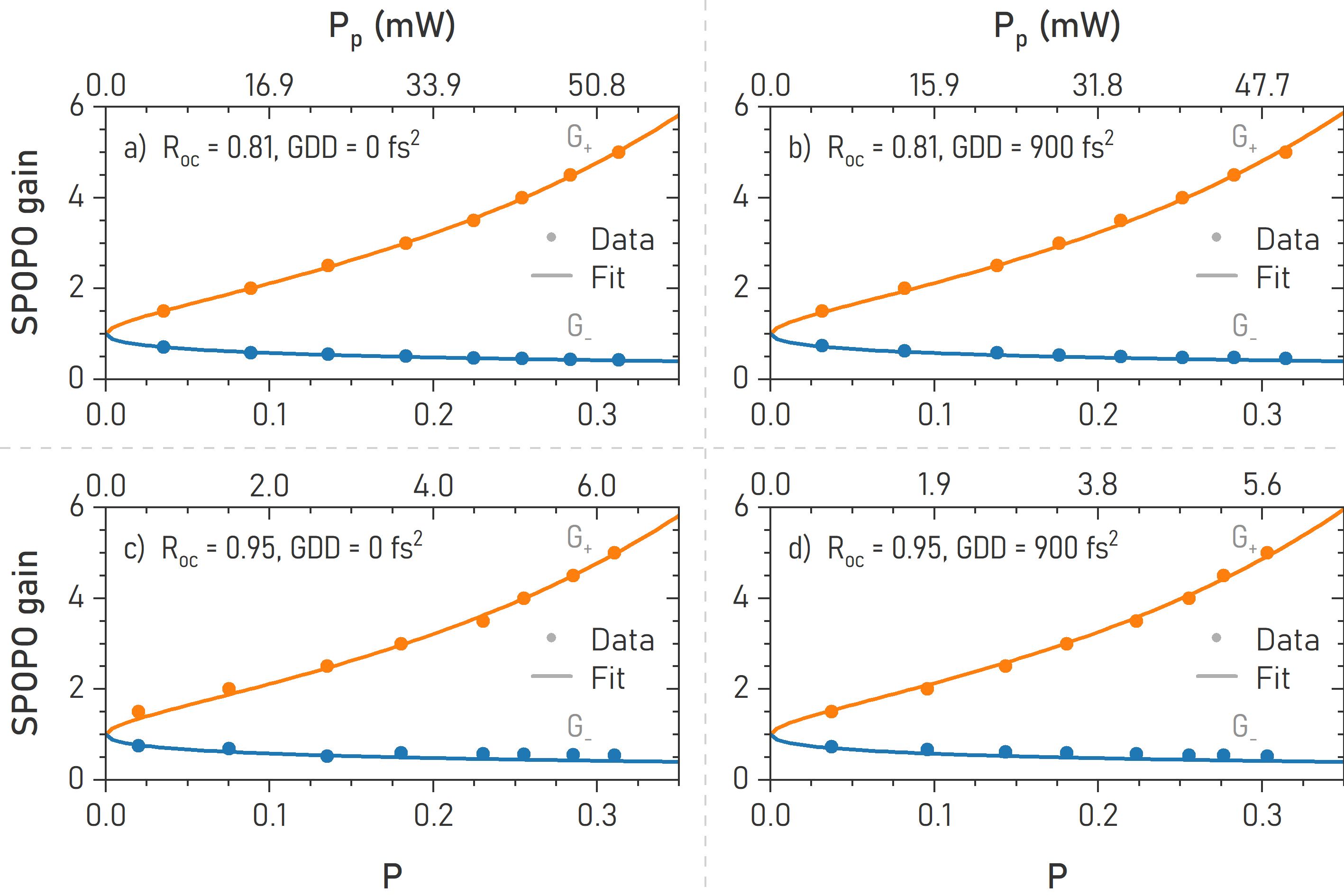}
    \caption{
    SPOPO parametric gain as a function of normalized pump power $P$ (bottom axis) and corresponding absolute pump power $P_p$ in \SI{}{\milli\watt} (top axis), for both amplification ($G_{+}$) and deamplification ($G_{-}$) regimes.  
    Data are reported for each cavity configuration. 
    }
    \label{fig:GDcal}
\end{figure}

\paragraph{SPOPO threshold calibration.}  
We begin the experimental characterization by calibrating the pump power relative to the SPOPO threshold.  
Hereafter, we denote the pump power in \SI{}{\milli\watt} as $P_p$, the SPOPO threshold power as $P_\mathrm{th}$, and the normalized pump power as $P = P_p / P_\mathrm{th}$.  
The normalized power $P$ is the parameter used in simulations, but only the absolute pump power $P_p$ is directly accessible, thus an accurate calibration of $P_\mathrm{th}$ is necessary.
To do so, the cavity was seeded with a weak coherent field instead of vacuum fluctuations (see AUX in \textbf{Figure}~\ref{fig:setup}). 
The seed beam, derived from the fundamental laser, was injected into the SPOPO and spatially and temporally overlapped with the pump field.  
By adjusting the relative phase between the seed and the pump, we recorded the two complementary responses: amplification of the seed ($G_{+}$) and deamplification of the seed ($G_{-}$).
We then measured $G_{+}$ and $G_{-}$ as functions of $P_p$. 
The experimental data were subsequently fitted to the theoretical expressions for the seeded SPOPO response, which predict~\cite{Roman-Rodriguez2024}  
\begin{align}\label{eq:gain}
    G_{\pm}(P) = \sum_k \frac{\left|N_k\right|^2}{\left(1 \mp \frac{\Delta'_k}{\Delta'_0}\sqrt{P}\right)^2} \, ,
\end{align}
where $N_k$ represents the projection of mode $k$ onto the weak injected seed, and $\Delta'_k$ are the eigenvalues of the interaction matrix describing the parametric process of Equation~\ref{eq:Lmatrixfil}, including the spectral filtering induced by dispersion. 
The resulting fits, displayed in \textbf{Figure}~\ref{fig:GDcal} alongside the measured points, demonstrate excellent agreement between theoretical predictions and experimental observations.  
This procedure allows the extraction of the threshold power $P_\mathrm{th}$ associated with the fundamental mode for each cavity configuration, providing a direct mapping between the experimentally accessible pump power $P_p$ and the normalized power $P$ employed throughout this work. 
Notice that the threshold power scales as $P_\mathrm{th} \propto \Delta_0^{-2}$~\cite{Patera2010}, where $\Delta_0$ is the eigenvalue of the fundamental supermode obtained from the SVD.
In the presence of intracavity dispersion, $\Delta_0^2$ decreases, so that one would expect a slightly higher threshold.
Experimentally, however, the observed trend is opposite.
The small discrepancy is compatible to experimental errors in power measurements and to slightly different alignments of the auxiliary beam (AUX) to the optical cavity in the two cases.

\begin{figure}
    \centering
    \includegraphics[width=120mm]{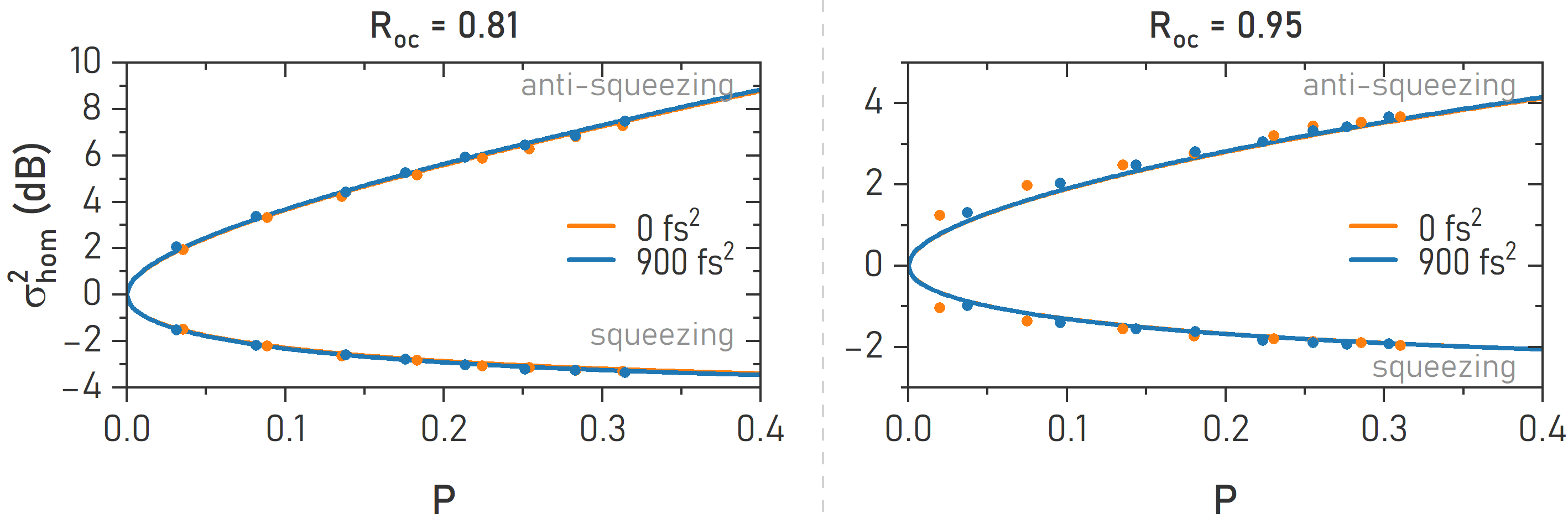}
    \caption{
    Squeezing and anti-squeezing as functions of the normalized pump power $P$, for both compensated (orange) and uncompensated (blue) dispersion.
    Left: $R_\mathrm{oc} = 0.81$. 
    Right: $R_\mathrm{oc} = 0.95$.
    In both cases $\Delta\lambda_\mathrm{LO}=\SI{2.7}{\nano\meter}$.
    Uncertainties are $\sim\pm\SI{0.2}{\decibel}$.
    }
    \label{fig:pump_dep}
\end{figure}

\paragraph{Squeezing versus pump power.}  
\textbf{Figure}~\ref{fig:pump_dep} presents the measured squeezing and anti-squeezing as functions of the normalized pump power $P$.  
Each panel compares the compensated (orange) and uncompensated (blue) cases for a given $R_\mathrm{oc}$. 
Error bars are within the symbol size in the plots.
The observed trends are essentially identical in both cases, for $R_\mathrm{oc}=0.81$ and $R_\mathrm{oc}=0.95$, with no significant differences in the measured squeezing levels. 
For $R_\mathrm{oc}=0.95$, the pump power $P_p$ of the first two points was low (below \SI{2}{\milli\watt}) and close to the detector resolution, leading to a slightly underestimated value of $P$ and a larger uncertainty.
On the other hand, this deviation is irrelevant for our analysis, as the comparison between the compensated and uncompensated cases remains clear.

\begin{figure}
    \centering
    \includegraphics[width=120mm]{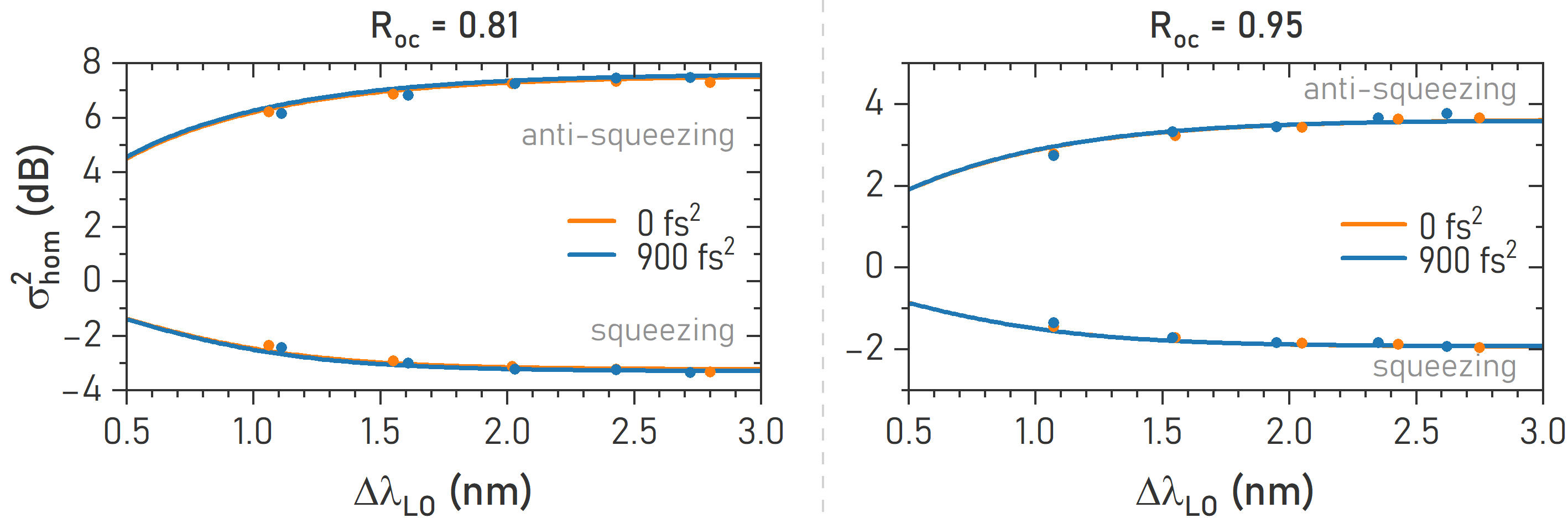}
    \caption{
    Squeezing and anti-squeezing as functions of the LO spectral bandwidth, for both compensated (orange) and uncompensated (blue) dispersion.
    Left: $R_\mathrm{oc} = 0.81$. 
    Right: $R_\mathrm{oc} = 0.95$.
    In both cases $P=0.31$.
    Uncertainties are $\sim\pm\SI{0.2}{\decibel}$.}
    \label{fig:lo_dep_sym}
\end{figure}

\begin{figure}
    \centering
    \includegraphics[width=120mm]{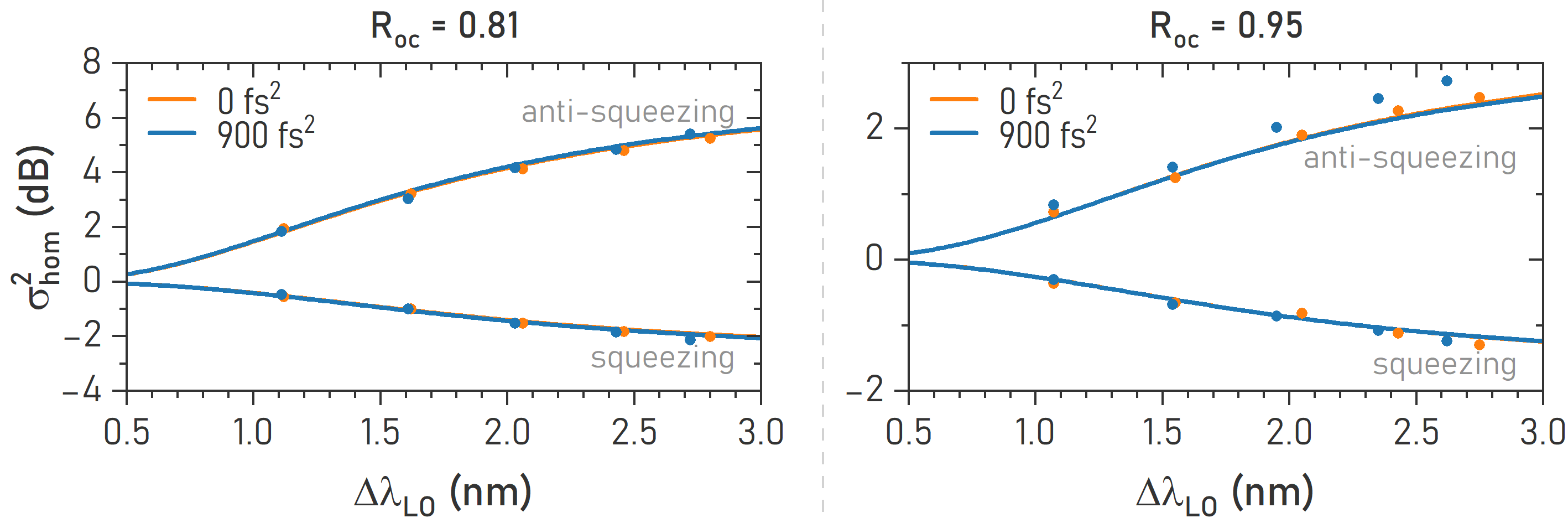}
    \caption{
    Squeezing and anti-squeezing as functions of the LO spectral bandwidth, for both compensated (orange) and uncompensated (blue) dispersion.
    Left: $R_\mathrm{oc} = 0.81$. 
    Right: $R_\mathrm{oc} = 0.95$.
    In both cases $P=0.31$.
    Uncertainties are $\sim\pm\SI{0.2}{\decibel}$.}
    \label{fig:lo_dep_asym}
\end{figure}

\paragraph{Squeezing versus LO bandwidth.}  
\textbf{Figure}~\ref{fig:lo_dep_sym} and \textbf{Figure}~\ref{fig:lo_dep_asym} show the measured squeezing and anti-squeezing as functions of the local oscillator (LO) spectral bandwidth, for both $R_\mathrm{oc}$ values and dispersion conditions. 
Error bars are within the symbol size in the plots.
Once again, the observed trends are nearly identical for compensated and uncompensated dispersion, and for both $R_\mathrm{oc}=0.81$ and $R_\mathrm{oc}=0.95$, with no appreciable difference in squeezing levels.
In the case of anti-symmetrical LO, the difference with respect to the zero dispersion case should have been significant.

\paragraph{Discussion}
Contrary to initial expectations, the experimental results indicate that the measured squeezing levels remain essentially unchanged when comparing compensated and uncompensated dispersion, both as a function of pump power and as a function of LO spectral bandwidth. 
This invariance suggests that the influence of intracavity dispersion is suppressed under our experimental conditions, resulting in a value of the parameter $K$ lower than expected.

As discussed in \textbf{Section}~\ref{sec:theory}, we interpret this behavior as resulting from the spectral filtering induced by intracavity dispersion on the generation of the SPOPO supermodes.
All simulations performed following this approach, i.e. starting from Equation~\ref{eq:Lmatrixfil}, and reported in the figures of this section, show excellent agreement with the experimental data, supporting the validity of this interpretation. 
Moreover, when including the effect of intracavity dispersion, we obtain $K=0.12$ for $R_\mathrm{oc} = 0.81$ and $\mathrm{GDD} = \SI{900}{\femto\second\squared}$, 
and $K = 0.21$ for $R_\mathrm{oc} = 0.95$ and $\mathrm{GDD} = \SI{900}{\femto\second\squared}$, i.e., nearly an order of magnitude smaller than expected, placing the system in a regime where dispersion-induced effects are low, in full agreement with the measurements.

Our experimental results indicate that, under realistic cavity parameters, intracavity dispersion may be far less detrimental to squeezing than previously anticipated, which has implications for the design of SPOPOs.
In particular, the reduction of the spectral bandwidth of the generated SPOPO field does not affect the measured squeezing level, as long as the LO spectrum is kept fixed.

\section{Conclusions}\label{sec:conclusions}
We have presented a detailed experimental study of dispersion effects in a SPOPO, combining precise pump calibration, spectral shaping of the local oscillator, and frequency-resolved homodyne detection. 
Across a range of pump powers, cavity finesses, and local oscillator bandwidths, the measured squeezing and anti-squeezing levels remain essentially unaffected by the presence or absence of intracavity group-delay dispersion. 
This invariance is at odds with simplified theoretical models, which predict a reduction in squeezing efficiency and a redistribution of energy into higher-order supermodes.
To account for this discrepancy, we propose a modeling approach in which the dispersion-induced spectral filtering is incorporated directly into the interaction Hamiltonian governing the generation of SPOPO supermodes, following the strategy illustrated in Ref.~\cite{Christ2014}.
Indeed, intracavity dispersion effectively acts as a spectral filter at the generation stage, narrowing the accessible mode spectrum and strongly reducing the sensitivity of the system to group-delay dispersion.
When implemented in this way, the model provides an excellent quantitative agreement with the experimental data across all investigated configurations, consistently reproducing the observed invariance of the squeezing level.

This interpretation extends existing theoretical descriptions and reconciles the apparent discrepancy between predictions and measurements, providing the first experimental benchmark that complements the theoretical framework of Ref.~\cite{Averchenko2024}. 

Future investigations could explore different dispersion regimes, nonlinear crystals, or cavity geometries to further clarify the interplay between dispersion, spectral filtering, and multimode squeezing, thereby strengthening the connection between theoretical predictions and experimental implementations in realistic SPOPO systems.

\appendix
\section{Dispersion spectral filtering}\label{appendix_A}
In this section, we derive the spectral shift of the cavity resonance induced by intracavity dispersion, leading to Equation~\ref{eq:dispshift}.
We consider a ring cavity of total length $L$, containing a dispersive nonlinear crystal of length $L_c$.
We denote by $\tilde{\omega}_j$ the angular frequency of the $j$-th cavity comb tooth, measured with respect to the central frequency $\tilde{\omega}_0$ corresponding to the index $j_0$.
Throughout this section, frequencies denoted by $\tilde{\omega}$ refer to the cavity comb, which is affected by dispersion, while $\omega$ refers to the frequencies of the radiation optical comb, in our case the quantum state.

After one round trip, the resonance condition for a cavity comb tooth at frequency $\tilde{\omega} = \tilde{\omega}_0 + \tilde{\omega}_j$ is given by a total phase $\tilde{\phi}_j= 2\pi\left(j_0 + j\right)$, i.e.,
\begin{align*}
    k\!\left(\tilde{\omega}_0 + \tilde{\omega}_j\right)\!\left(L-L_c\right) +
    k_m\!\left(\tilde{\omega}_0 + \tilde{\omega}_j\right)\!L_c =
    2\pi\left(j_0 + j\right) \, ,
\end{align*}
where $k$ is the free-space wavenumber (assumed constant) and $k_m$ is the wavenumber inside the nonlinear medium.
Expanding $k_m$ up to the second order around $\tilde{\omega}_0$, we obtain
\begin{align*}
    &k\!\left(\tilde{\omega}_0+\tilde{\omega}_j\right)L - 
    k\!\left(\tilde{\omega}_0+\tilde{\omega}_j\right)L_c + \\
    & + k_m\!\left(\tilde{\omega}_0\right)L_c + 
    \left.\frac{\partial k_m}{\partial {\omega}}\right|_{\tilde{\omega}_0}\! \tilde{\omega}_j L_c +
    \frac12\left.\frac{\partial^2 k_m}{\partial {\omega}^2}\right|_{\tilde{\omega}_0}\! \tilde{\omega}_j^2 L_c =
    2\pi\left(j_0 + j\right) \, .
\end{align*}

Since $k = n\,\omega/c$, $\frac{\partial k_m}{\partial \omega} = n_{g}/c$ (where $n_g$ is the group refractive index), and $\frac{\partial^2 k_m}{\partial \omega^2} = D$, defining $n_{m,g}$ as the group refractive index in the nonlinear medium, we can rewrite
\begin{align*}
    &\frac{n_0}{c} \left(\tilde{\omega}_0 + \tilde{\omega}_j\right)L -
    \frac{n_0}{c} \left(\tilde{\omega}_0 + \tilde{\omega}_j\right)L_c + \\
    &+ \frac{n_{m,0}}{c}\tilde{\omega}_0 L_c + 
    \frac{n_{m,g}}{c}\tilde{\omega}_j L_c +
    \frac12 D\,L_c \tilde{\omega}_j^2 = 
    2\pi\left(j_0 + j\right) \, .
\end{align*}

The resonance condition can thus be cast as
\begin{align*}
    \frac{\tilde{\omega}_0}{c}L_\mathrm{eff} +
    \frac{\tilde{\omega}_j}{c}L_g +
    \frac12 \mathrm{GDD} \, \tilde{\omega}_j^2 =
    2\pi\left(j_0 + j\right) \, ,
\end{align*}
where we defined $L_\mathrm{eff} = L+L_c\left(n_{m,0}-1\right)$, $L_g = L+L_c\left(n_{m,g}-1\right)$, and $D\,L_c = \mathrm{GDD}$.
It follows that
\begin{align}\label{eq:omegaj1}
    &\frac{\tilde{\omega}_0}{c}L_\mathrm{eff} = 2\pi j_0
\end{align}
and
\begin{align}\label{eq:omegaj2}
    \frac{\tilde{\omega}_j}{c}L_g + \frac12 \mathrm{GDD}\,\tilde{\omega}_j^2 = 2\pi j \, .
\end{align}
Equation \ref{eq:omegaj1} defines the resonance condition for the reference (central) cavity comb tooth, while Equation~\ref{eq:omegaj2} determines the frequency of the $j$th comb tooth in the presence of dispersion.
Solving Equation~\ref{eq:omegaj2} for $\tilde{\omega}_j$ yields
\begin{align*}
    \tilde{\omega}_j = \frac{1}{\mathrm{GDD}}
    \left(
    -\frac{L_g}{c} \pm \sqrt{\left(\frac{L_g}{c}\right)^2 + 4\pi\,\mathrm{GDD}\,j}
    \right) \, .
\end{align*}
We retain the positive root, since $\tilde{\omega}_j$ increases with $j$.
Expanding the square root to second order and defining $\frac{L_g}{c} = \mathrm{FSR}^{-1}$ \cite{Thorpe2008}, and $\omega_j = 2\pi j\,\mathrm{FSR}$ as the frequency of the $j$th tooth of the optical radiation comb, we obtain
\begin{align}\label{eq:toothfreq}
    \tilde{\omega}_j = 2\pi j\,\mathrm{FSR} +
    \frac{1}{2}\,\mathrm{GDD}\,\mathrm{FSR}\,\left(2\pi j\,\mathrm{FSR}\right)^2
    = \omega_j + \frac{1}{2}\,\mathrm{GDD}\,\mathrm{FSR}\,\omega_j^2 \, .
\end{align}
Therefore, the detuning between the cavity comb tooth and the corresponding optical radiation tooth is
\begin{align*}
    \delta\omega_j = \tilde{\omega}_j - \omega_j = \frac{1}{2}\,\mathrm{GDD}\,\mathrm{FSR}\,\omega_j^2 \, ,
\end{align*}
as reported in Equation~\ref{eq:dispshift}.
The equation shows that intracavity GDD induces a quadratic detuning of the cavity resonance frequencies, causing a systematic mismatch between the cavity and the optical comb modes as the frequency departs from the carrier. 
This effect plays a central role in determining the phase matching and the effective gain bandwidth in cavity-enhanced parametric processes.

\medskip
\textbf{Data Availability Statement} \par
The datasets generated and analyzed during the current study are available from the corresponding author on reasonable request.

\medskip
\textbf{Funding} \par
This work has been funded by the Istituto Nazionale di Fisica Nucleare (INFN) within the project T4QC. 
Sebastiano Corli has been supported by the project QXtreme.

\medskip
\textbf{Acknowledgements} \par 
The authors acknowledge Ennio Viscione of Servizio Progettazione Meccanica e Officina at INFN LASA.

\medskip

%
\bibliography{biblio}

@article{Asavanant2019,
author = {Asavanant, Warit and Shiozawa, Yu and Yokoyama, Shota and Charoensombutamon, Baramee and Emura, Hiroki and Alexander, Rafael N. and Takeda, Shuntaro and Yoshikawa, Jun-ichi and Menicucci, Nicolas C. and Yonezawa, Hidehiro and Furusawa, Akira},
doi = {10.1126/science.aay2645},
issn = {0036-8075},
journal = {Science},
month = {oct},
number = {6463},
pages = {373--376},
title = {{Generation of time-domain-multiplexed two-dimensional cluster state}},
url = {https://www.science.org/doi/10.1126/science.aay2645},
volume = {366},
year = {2019}
}

@article{Averchenko2024,
archivePrefix = {arXiv},
arxivId = {2407.18826},
author = {Averchenko, V A and Malyshev, D M and Tikhonov, K S},
doi = {10.1088/1367-2630/ad9be1},
eprint = {2407.18826},
issn = {1367-2630},
journal = {New Journal of Physics},
month = {dec},
number = {12},
pages = {123017},
title = {{Effect of group-velocity dispersion on the generation of multimode pulsed squeezed light in a synchronously pumped optical parametric oscillator}},
url = {http://arxiv.org/abs/2407.18826 https://iopscience.iop.org/article/10.1088/1367-2630/ad9be1},
volume = {26},
year = {2024}
}

@book{Bachor2019,
author = {Bachor, Hans‐A. and Ralph, Timothy C.},
doi = {10.1002/9783527695805},
isbn = {9783527411931},
month = {sep},
publisher = {Wiley},
title = {{A Guide to Experiments in Quantum Optics}},
url = {https://onlinelibrary.wiley.com/doi/book/10.1002/9783527695805},
year = {2019}
}

@book{ParisJarda2004,
editor = {Paris, Matteo G. A. and Řeháček, Jaroslav},
title = {Quantum State Estimation},
publisher = {Springer},
address ={Berlin},
year = {2004},
series = {LNP 649}
}

@article{Braunstein2005,
author = {Braunstein, Samuel L.},
doi = {10.1103/PhysRevA.71.055801},
issn = {1050-2947},
journal = {Physical Review A},
month = {may},
number = {5},
pages = {055801},
title = {{Squeezing as an irreducible resource}},
url = {https://link.aps.org/doi/10.1103/PhysRevA.71.055801},
volume = {71},
year = {2005}
}

@article{Conlon2024,
author = {Conlon, Lorc{\'{a}}n O. and Shajilal, Biveen and Walsh, Angus and Zhao, Jie and Janousek, Jiri and Lam, Ping Koy and Assad, Syed M.},
doi = {10.1038/s41534-024-00834-9},
issn = {2056-6387},
journal = {npj Quantum Information},
month = {apr},
number = {1},
pages = {35},
title = {{Verifying the security of a continuous variable quantum communication protocol via quantum metrology}},
url = {https://www.nature.com/articles/s41534-024-00834-9},
volume = {10},
year = {2024}
}

@article{Crisafulli2013,
author = {Crisafulli, Orion and Tezak, Nikolas and Soh, Daniel B. S. and Armen, Michael A. and Mabuchi, Hideo},
doi = {10.1364/OE.21.018371},
issn = {1094-4087},
journal = {Optics Express},
month = {jul},
number = {15},
pages = {18371},
title = {{Squeezed light in an optical parametric oscillator network with coherent feedback quantum control}},
url = {https://opg.optica.org/oe/abstract.cfm?uri=oe-21-15-18371},
volume = {21},
year = {2013}
}

@article{Drever1983,
author = {Drever, R. W. P. and Hall, J. L. and Kowalski, F. V. and Hough, J. and Ford, G. M. and Munley, A. J. and Ward, H.},
doi = {10.1007/BF00702605},
issn = {0721-7269},
journal = {Applied Physics B Photophysics and Laser Chemistry},
month = {jun},
number = {2},
pages = {97--105},
title = {{Laser phase and frequency stabilization using an optical resonator}},
url = {http://link.springer.com/10.1007/BF00702605},
volume = {31},
year = {1983}
}

@article{Fabre2020,
author = {Fabre, C. and Treps, N.},
doi = {10.1103/RevModPhys.92.035005},
issn = {0034-6861},
journal = {Reviews of Modern Physics},
month = {sep},
number = {3},
pages = {035005},
title = {{Modes and states in quantum optics}},
url = {https://link.aps.org/doi/10.1103/RevModPhys.92.035005},
volume = {92},
year = {2020}
}

@article{Grice2001,
author = {Grice, W. P. and U'Ren, A. B. and Walmsley, I. A.},
doi = {10.1103/PhysRevA.64.063815},
issn = {1050-2947},
journal = {Physical Review A},
month = {nov},
number = {6},
pages = {063815},
title = {{Eliminating frequency and space-time correlations in multiphoton states}},
url = {https://link.aps.org/doi/10.1103/PhysRevA.64.063815},
volume = {64},
year = {2001}
}

@article{Jiang2012,
author = {Jiang, Shifeng and Treps, Nicolas and Fabre, Claude},
doi = {10.1088/1367-2630/14/4/043006},
issn = {13672630},
journal = {New Journal of Physics},
title = {{A time/frequency quantum analysis of the light generated by synchronously pumped optical parametric oscillators}},
volume = {14},
year = {2012}
}

@article{Kaiser2016,
author = {Kaiser, F. and Fedrici, B. and Zavatta, A. and D'Auria, V. and Tanzilli, S.},
doi = {10.1364/OPTICA.3.000362},
issn = {2334-2536},
journal = {Optica},
month = {apr},
number = {4},
pages = {362},
title = {{A fully guided-wave squeezing experiment for fiber quantum networks}},
url = {https://opg.optica.org/abstract.cfm?URI=optica-3-4-362},
volume = {3},
year = {2016}
}

@article{Kamble2024,
author = {Kamble, Mrunal and Wang, Jiaxuan and Agarwal, Girish S.},
doi = {10.1103/PhysRevA.109.053715},
issn = {2469-9926},
journal = {Physical Review A},
month = {may},
number = {5},
pages = {053715},
title = {{Quantum metrology of absorption and gain parameters using two-mode bright squeezed light}},
url = {https://link.aps.org/doi/10.1103/PhysRevA.109.053715},
volume = {109},
year = {2024}
}

@article{Larsen2019,
author = {Larsen, Mikkel V. and Guo, Xueshi and Breum, Casper R. and Neergaard-Nielsen, Jonas S. and Andersen, Ulrik L.},
doi = {10.1126/science.aay4354},
issn = {0036-8075},
journal = {Science},
month = {oct},
number = {6463},
pages = {369--372},
title = {{Deterministic generation of a two-dimensional cluster state}},
url = {https://www.science.org/doi/10.1126/science.aay4354},
volume = {366},
year = {2019}
}

@article{Lawrie2019,
author = {Lawrie, B. J. and Lett, P. D. and Marino, A. M. and Pooser, R. C.},
doi = {10.1021/acsphotonics.9b00250},
issn = {2330-4022},
journal = {ACS Photonics},
month = {jun},
number = {6},
pages = {1307--1318},
title = {{Quantum Sensing with Squeezed Light}},
url = {https://pubs.acs.org/doi/10.1021/acsphotonics.9b00250},
volume = {6},
year = {2019}
}

@article{Leonhardt1995,
author = {Leonhardt, U. and Paul, H.},
doi = {10.1016/0079-6727(94)00007-L},
issn = {00796727},
journal = {Progress in Quantum Electronics},
month = {jan},
number = {2},
pages = {89--130},
title = {{Measuring the quantum state of light}},
url = {https://linkinghub.elsevier.com/retrieve/pii/007967279400007L},
volume = {19},
year = {1995}
}

@article{Madsen2022,
author = {Madsen, Lars S. and Laudenbach, Fabian and Askarani, Mohsen Falamarzi. and Rortais, Fabien and Vincent, Trevor and Bulmer, Jacob F. F. and Miatto, Filippo M. and Neuhaus, Leonhard and Helt, Lukas G. and Collins, Matthew J. and Lita, Adriana E. and Gerrits, Thomas and Nam, Sae Woo and Vaidya, Varun D. and Menotti, Matteo and Dhand, Ish and Vernon, Zachary and Quesada, Nicol{\'{a}}s and Lavoie, Jonathan},
doi = {10.1038/s41586-022-04725-x},
issn = {0028-0836},
journal = {Nature},
month = {jun},
number = {7912},
pages = {75--81},
title = {{Quantum computational advantage with a programmable photonic processor}},
url = {https://www.nature.com/articles/s41586-022-04725-x},
volume = {606},
year = {2022}
}

@article{Menicucci2014,
author = {Menicucci, Nicolas C.},
doi = {10.1103/PhysRevLett.112.120504},
issn = {0031-9007},
journal = {Physical Review Letters},
month = {mar},
number = {12},
pages = {120504},
title = {{Fault-Tolerant Measurement-Based Quantum Computing with Continuous-Variable Cluster States}},
url = {https://link.aps.org/doi/10.1103/PhysRevLett.112.120504},
volume = {112},
year = {2014}
}

@article{Monmayrant2010,
author = {Monmayrant, Antoine and Weber, S{\'{e}}bastien and Chatel, B{\'{e}}atrice},
doi = {10.1088/0953-4075/43/10/103001},
issn = {0953-4075},
journal = {Journal of Physics B: Atomic, Molecular and Optical Physics},
month = {may},
number = {10},
pages = {103001},
title = {{A newcomer's guide to ultrashort pulse shaping and characterization}},
url = {https://iopscience.iop.org/article/10.1088/0953-4075/43/10/103001},
volume = {43},
year = {2010}
}

@article{Mosley2008,
author = {Mosley, Peter J. and Lundeen, Jeff S. and Smith, Brian J. and Wasylczyk, Piotr and U'Ren, Alfred B. and Silberhorn, Christine and Walmsley, Ian A.},
doi = {10.1103/PhysRevLett.100.133601},
issn = {0031-9007},
journal = {Physical Review Letters},
month = {apr},
number = {13},
pages = {133601},
title = {{Heralded Generation of Ultrafast Single Photons in Pure Quantum States}},
url = {https://link.aps.org/doi/10.1103/PhysRevLett.100.133601},
volume = {100},
year = {2008}
}

@article{Patera2010,
author = {Patera, G. and Treps, N. and Fabre, C. and de Valc{\'{a}}rcel, G. J.},
doi = {10.1140/epjd/e2009-00299-9},
issn = {1434-6060},
journal = {The European Physical Journal D},
month = {jan},
number = {1},
pages = {123--140},
title = {{Quantum theory of synchronously pumped type I optical parametric oscillators: characterization of the squeezed supermodes}},
url = {http://link.springer.com/10.1140/epjd/e2009-00299-9},
volume = {56},
year = {2010}
}

@article{Plick2018,
author = {Plick, William N. and Arzani, Francesco and Treps, Nicolas and Diamanti, Eleni and Markham, Damian},
doi = {10.1103/PhysRevA.98.062101},
issn = {2469-9926},
journal = {Physical Review A},
month = {dec},
number = {6},
pages = {062101},
title = {{Violating Bell inequalities with entangled optical frequency combs and multipixel homodyne detection}},
url = {https://link.aps.org/doi/10.1103/PhysRevA.98.062101},
volume = {98},
year = {2018}
}

@article{Roman-Rodriguez2024,
author = {Roman-Rodriguez, Victor and Fainsin, David and Zanin, Guilherme L. and Treps, Nicolas and Diamanti, Eleni and Parigi, Valentina},
doi = {10.1103/PhysRevResearch.6.043113},
issn = {2643-1564},
journal = {Physical Review Research},
month = {nov},
number = {4},
pages = {043113},
title = {{Multimode squeezed state for reconfigurable quantum networks at telecommunication wavelengths}},
url = {https://link.aps.org/doi/10.1103/PhysRevResearch.6.043113},
volume = {6},
year = {2024}
}

@article{Roslund2014,
author = {Roslund, Jonathan and de Ara{\'{u}}jo, Renn{\'{e}} Medeiros and Jiang, Shifeng and Fabre, Claude and Treps, Nicolas},
doi = {10.1038/nphoton.2013.340},
issn = {1749-4885},
journal = {Nature Photonics},
month = {feb},
number = {2},
pages = {109--112},
title = {{Wavelength-multiplexed quantum networks with ultrafast frequency combs}},
url = {https://www.nature.com/articles/nphoton.2013.340},
volume = {8},
year = {2014}
}

@article{Suerra2025,
author = {Suerra, Edoardo and Canella, Francesco and Giannotti, Dario and Khalili, Mohsen and Wang, Yicheng and Hasse, Kore and Suntsov, Sergiy and Kip, Detlef and Saraceno, Clara and Cialdi, Simone and Galzerano, Gianluca},
doi = {10.1063/5.0252040},
issn = {2378-0967},
journal = {APL Photonics},
month = {apr},
number = {4},
title = {{Ytterbium-laser-driven THz generation in thin lithium niobate at 1.9 kW average power in a passive enhancement cavity}},
url = {https://pubs.aip.org/app/article/10/4/046111/3344233/Ytterbium-laser-driven-THz-generation-in-thin},
volume = {10},
year = {2025}
}

@article{Thorpe2008,
author = {Thorpe, M.J. and Ye, J.},
doi = {10.1007/s00340-008-3019-1},
issn = {0946-2171},
journal = {Applied Physics B},
month = {jun},
number = {3-4},
pages = {397--414},
title = {{Cavity-enhanced direct frequency comb spectroscopy}},
url = {http://link.springer.com/10.1007/s00340-008-3019-1},
volume = {91},
year = {2008}
}

@article{Walmsley2023,
author = {Walmsley, Ian},
doi = {10.1364/OPTICAQ.507527},
issn = {2837-6714},
journal = {Optica Quantum},
month = {oct},
number = {1},
pages = {35},
title = {{Light in quantum computing and simulation: perspective}},
url = {https://opg.optica.org/abstract.cfm?URI=opticaq-1-1-35},
volume = {1},
year = {2023}
}

@inproceedings{Walschaers2019,
address = {Washington, D.C.},
author = {Walschaers, Mattia and Parigi, Valentina and Treps, Nicolas},
booktitle = {Quantum Information and Measurement (QIM) V: Quantum Technologies},
doi = {10.1364/QIM.2019.S2B.5},
isbn = {978-1-943580-56-9},
pages = {S2B.5},
publisher = {OSA},
title = {{Photon-Subtracted Continuous-Variable Graph States}},
url = {https://opg.optica.org/abstract.cfm?URI=QIM-2019-S2B.5},
year = {2019}
}

@article{Walschaers2023,
author = {Walschaers, Mattia and Sundar, Bhuvanesh and Treps, Nicolas and Carr, Lincoln D and Parigi, Valentina},
doi = {10.1088/2058-9565/accdfd},
issn = {2058-9565},
journal = {Quantum Science and Technology},
month = {jul},
number = {3},
pages = {035009},
title = {{Emergent complex quantum networks in continuous-variables non-Gaussian states}},
url = {https://iopscience.iop.org/article/10.1088/2058-9565/accdfd},
volume = {8},
year = {2023}
}

@article{Zhong2021,
author = {Zhong, Han-Sen and Deng, Yu-Hao and Qin, Jian and Wang, Hui and Chen, Ming-Cheng and Peng, Li-Chao and Luo, Yi-Han and Wu, Dian and Gong, Si-Qiu and Su, Hao and Hu, Yi and Hu, Peng and Yang, Xiao-Yan and Zhang, Wei-Jun and Li, Hao and Li, Yuxuan and Jiang, Xiao and Gan, Lin and Yang, Guangwen and You, Lixing and Wang, Zhen and Li, Li and Liu, Nai-Le and Renema, Jelmer J. and Lu, Chao-Yang and Pan, Jian-Wei},
doi = {10.1103/PhysRevLett.127.180502},
issn = {0031-9007},
journal = {Physical Review Letters},
month = {oct},
number = {18},
pages = {180502},
title = {{Phase-Programmable Gaussian Boson Sampling Using Stimulated Squeezed Light}},
url = {https://link.aps.org/doi/10.1103/PhysRevLett.127.180502},
volume = {127},
year = {2021}
}

@article{Suerra2026,
author = {Suerra, Edoardo and Altilia, Samuele and Olivares, Stefano and Ferraro, Alessandro and Canella, Francesco and Giannotti, Dario and Galzerano, Gianluca and Corli, Sebastiano and Prati, Enrico and Cialdi, Simone},
doi = {10.1007/s00340-025-08617-6},
issn = {1432-0649},
journal = {Applied Physics B},
month = {feb},
number = {3},
pages = {26},
title = {{Generation and detection of squeezed states via a synchronously pumped optical parametric oscillator}},
url = {https://link.springer.com/10.1007/s00340-025-08617-6},
volume = {132},
year = {2026}
}

@incollection{DARIANO2003,
title = {Quantum Tomography},
editor = {Peter W. Hawkes},
series = {Advances in Imaging and Electron Physics},
publisher = {Elsevier},
volume = {128},
pages = {205-308},
year = {2003},
issn = {1076-5670},
doi = {https://doi.org/10.1016/S1076-5670(03)80065-4},
url = {https://www.sciencedirect.com/science/article/pii/S1076567003800654},
author = {G. {Mauro D’Ariano} and Matteo G.A. Paris and Massimiliano F. Sacchi}
}

@article{Olivares21,
title = {Introduction to generation, manipulation and characterization of optical quantum states},
journal = {Physics Letters A},
volume = {418},
pages = {127720},
year = {2021},
issn = {0375-9601},
doi = {https://doi.org/10.1016/j.physleta.2021.127720},
url = {https://www.sciencedirect.com/science/article/pii/S0375960121005843},
author = {Stefano Olivares}
}

@book{Siegman1986,
address = {Mill Valley, CA, USA},
author = {{Anthony E. Siegman}},
isbn = {9780935702118},
publisher = {University Science Books},
title = {{Lasers}},
year = {1986}
}

@book{Svelto2010,
address = {New York, NY},
author = {Svelto, Orazio},
booktitle = {Principles of Lasers},
doi = {10.1007/978-1-4419-1302-9},
edition = {5},
isbn = {978-1-4419-1302-9},
issn = {0010-7514},
pages = {XXI, 620},
publisher = {Springer US},
title = {{Principles of Lasers}},
url = {http://link.springer.com/10.1007/978-1-4615-7670-9 http://www.tandfonline.com/doi/abs/10.1080/00107514.2011.647714 http://link.springer.com/10.1007/978-1-4419-1302-9},
year = {2010}
}

@article{Christ2014,
author = {Christ, Andreas and Lupo, Cosmo and Reichelt, Matthias and Meier, Torsten and Silberhorn, Christine},
doi = {10.1103/PhysRevA.90.023823},
issn = {1050-2947},
journal = {Physical Review A},
month = {aug},
number = {2},
pages = {023823},
title = {{Theory of filtered type-II parametric down-conversion in the continuous-variable domain: Quantifying the impacts of filtering}},
url = {https://link.aps.org/doi/10.1103/PhysRevA.90.023823},
volume = {90},
year = {2014}
}

\end{document}